\documentclass{llncs}
\usepackage{graphicx}
\usepackage{natbib}
\usepackage{multirow}
\usepackage{rotating}
\usepackage[top=1in, bottom=1in, left=1in, right=1in]{geometry}

\begin{document}
\frontmatter          
\pagestyle{headings}  

\title{Time-Varying Networks: Recovering Temporally Rewiring Genetic Networks During the Life Cycle
of \\
\textit{Drosophila melanogaster}}%

\titlerunning{}  

\authorrunning{}   
%
\author{Amr Ahmed$^\dagger$, Le Song$^\dagger$ and Eric P. Xing$^\star$}
\institute{ SAILING Lab, Carnegie Mellon
University\\
\{amahmed,lesong,epxing\}@cs.cmu.edu\\
$^\dagger$equally contributing authors\\
$^\star$ corresponding author}


\maketitle              

\begin{abstract}

Due to the dynamic nature of biological systems, biological networks
underlying temporal process such as the development of {\it
Drosophila melanogaster} can exhibit significant topological changes
to facilitate dynamic regulatory functions. Thus it is essential to
develop methodologies that capture the temporal evolution of
networks, which make it possible to study the driving forces
underlying dynamic rewiring of gene regulation circuity, and to
predict future network structures. Using a new machine learning
method called Tesla, which builds on a novel temporal logistic
regression technique, we report the first successful genome-wide
reverse-engineering of the latent sequence of temporally rewiring
gene networks over more than 4000 genes during the life cycle of
\textit{Drosophila melanogaster}, given longitudinal gene expression
measurements and even when a single snapshot of such measurement
resulted from each (time-specific) network is available. Our methods
offer the first glimpse of time-specific snapshots and temporal
evolution patterns of gene networks in a living organism during its
full developmental course. The recovered networks with this
unprecedented resolution chart the onset and duration of many gene
interactions which are missed by typical static network analysis,
and are suggestive of a wide array of other temporal behaviors of
the gene network over time not noticed before.

\end{abstract}

\newpage

\section*{Introduction}

A major challenge in systems biology is to understand and model,
quantitatively, the topological, functional, and dynamical
properties of cellular networks, such as transcriptional regulatory
circuitry and signal transaction pathways, that control the
behaviors of the cell.

Empirical studies showed that many biological networks bear
remarkable similarities to various other networks in nature, such as
social networks, in terms of global topological characteristics such
as the scale-free and small-world properties, albeit with different
characteristic coefficients~\citep{Bara:Albe:1999}. Furthermore, it
was observed that the average clustering factor of real biological
networks is significantly larger than that of a random network of
equivalent size and degree distribution~\citep{Bara:Oltv:2004}; and
biological networks are characterized by their intrinsic
modularities~\citep{Vasz:Dobr:Serg:Eckm:2004}, which reflect
presence of physically and/or functionally linked molecules that
work synergistically to achieve a relatively autonomous
functionality. These studies have led to numerous advances towards
uncovering the organizational principles and functional properties
of biological networks, and even identification of new regulatory
events~\citep{Califano}; however, most of such results are based on
analyses of {\it static networks}, that is, networks with invariant
topology over a given set of molecules, such as a protein-protein
interaction (PPI) network over all proteins of an organism
regardless of the conditions under which individual interactions may
take place, or a single gene network inferred from microarray data
even though the samples may be collected over a time course or
multiple conditions.

Over the course of a cellular process, such as a cell cycle or an
immune response, there may exist multiple underlying "themes" that
determine the functionalities of each molecule and their
relationships to each other, and such themes are dynamical and
stochastic. As a result, the molecular networks at each time point
are context-dependent and can undergo systematic rewiring, rather
than being invariant over time, as assumed in most current
biological network studies. Indeed, in a seminal study
by~\cite{gerstein}, it was shown that the "active regulatory paths"
in a gene expression correlation network of {\em Saccharomyces
cerevisiae} exhibit dramatic topological changes and hub transience
during a temporal cellular process, or in response to diverse
stimuli. However, the exact mechanisms underlying this phenomena
remain poorly understood.

What prevents us from an in-depth investigation of the mechanisms that
drive the temporal rewiring of biological networks during various cellular and
physiological processes? A key technical hurdle we face  is the unavailability of {\em serial snapshots}
of the rewiring network during a biological process. 
Under a realistic dynamic biological system, usually it is
technologically impossible to experimentally determine time-specific
network topologies for a series of time points based on techniques
such as two-hybrid or ChIP-chip systems; resorting to computational
inference methods such as structural learning algorithms for
Bayesian networks is also difficult because we can only obtain a
single snapshot of the gene expressions at each time point -- how
can one derive a network structure specific to a point of time based
on only one measurement of node-states at that time? If we follow
the naive assumption that each snapshot is from a different network,
this task would be statistically impossible because our estimator
(from only one sample) would suffer from extremely high variance.
Extant methods would instead pool samples from all time points
together and infer a single "average"
network~\citep{Friedman2000,ong,Califano}, which means they choose
to ignore network rewiring and simply assume that the network
snapshots are independently and identically distributed. Or one
could perhaps divide the time series into overlapping sliding
windows and infer static networks for each window separately. These
approach, however, only utilizes a limited number of samples in each
window and ignores the smoothly evolving nature of the networks.
Therefore, the resulting networks are limited in term of their
temporal resolution and statistical power. To our knowledge, no
method is currently available for genome-wide reverse engineering of
time-varying networks underlying biological processes with temporal
resolution up to every single time point where gene expressions are
measured.

Here, we present the first successful reverse engineering of a
series of 23 time-varying networks of {\it Drosophila melanogaster}
over its entire developmental course, and a detailed topological
analysis of the temporal evolution patterns in this series of
networks. Our study is based on a new machine learning algorithm
called  {\it TEmporally Smoothed $L_1$-regularized LOgistic
Regression}, or Tesla (stemmed from TESLLOR, the acronym of our
algorithm). Tesla is based on a key assumption that temporally
adjacent networks are likely not to be dramatically different from
each other in topology, and therefore are more likely to share
common edges than temporally distal networks. Building on the
powerful and highly scalable iterative $L_1$-regularized logistic
regression algorithm for estimating single sparse
networks~\citep{Martin}, we develop a novel regression
regularization scheme that connects multiple time-specific network
inference functions via a first-order edge smoothness function that
encourages edge retention in networks immediately across time
points.  An important property of this novel idea is that it fully
integrates all available samples of the entire time series in a
single inference procedure that recovers the wiring patterns between
genes over a time series of arbitrary resolution --- from a network
for every single time point, to one network for every $K$ time
points where $K$ is very small. Besides, our method can also benefit
from the smoothly evolving nature of the underlying networks and the
prior knowledge on gene ontology groups. These additional pieces of
information increase the chance to recover biologically plausible
networks while at the same time reduce the computational complexity
of the inference algorithms. Importantly, Tesla can be casted as a
convex optimization problem for which a globally optimal solution
exists and can be efficiently computed for networks with thousands
of nodes.

To our knowledge, Tesla represents the first successful attempt on
genome-wide reverse engineering of time-varying networks underlying
biological processes with arbitrary temporal resolution. Earlier
algorithmic approaches, such as the structure learning algorithms
for dynamic Bayesian network~\citep{ong}, learns a time-homogeneous
dynamic system with fixed node dependencies, which is entirely
different from our goal, which aims at snapshots of rewiring
network. The Trace-back algorithm~\citep{gerstein} that enables the
revelation of network changes over time in yeast is based on tracing
active paths or subnetwork in static summary network estimated {\it
a priori} from all samples from a time series, which is
significantly different from our method, because edges that are
transient over a short period of time may be missed by the summary
network in the first place. The DREM program~\citep{Ernst-Ziv} that
reconstructs dynamic regulatory maps tracks bifurcation points of a
regulatory cascade according to the ChIP-chip data over short time
course, which is also different from our method, because Tesla aims
at recover the entire time-varying networks, not only the
interactions due to protein-DNA binding, from long time series with
arbitrary temporal resolution.

A recent genome-wide microarray profiling of the life cycle of
\textit{Drosophila melanogaster} revealed the evolving nature of the
gene expression patterns during the time course of its
development~\citep{Arbeitmanetal2002}. In this study, 4028 genes
were examined at 66 distinct time points spanning the embryonic,
larval, pupal and adulthood period of the organism. It was found
that most genes ($86\%$) were differentially expressed over time,
and the time courses of many genes followed a wave structure with
one to three peaks. Furthermore, clustering genes according their
expression profiles helped identify functional coherent groups
specific to tissue and organ development. Using Tesla, we
successfully reverse-engineered a sequence of 23 epoch-specific
networks from this data set. Detailed analysis of these networks
reveals a rich set of information and findings regarding how crucial
network statistics change over time, the dynamic behavior of key
genes at the hubs of the networks, how genes forms dynamic clusters
and how their inter-connectivity rewires overtime, and improved
prediction of the activation patterns of already known gene
interactions and link them to functional behaviors and developmental
stages.

\section*{Results and Discussion}

\subsection*{Time-Varying Gene Networks in \textit{Drosophila
melanogaster}}

\begin{sidewaysfigure}
    \begin{tabular}{|c|c|c|c|c|c||c|}
    \hline
    \hline
    &&&&&&
    \multirow{5}{*}{\includegraphics[width=0.28\columnwidth]{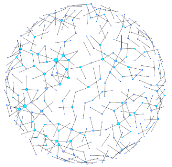}}
    \\
    \includegraphics[width=0.09\columnwidth]{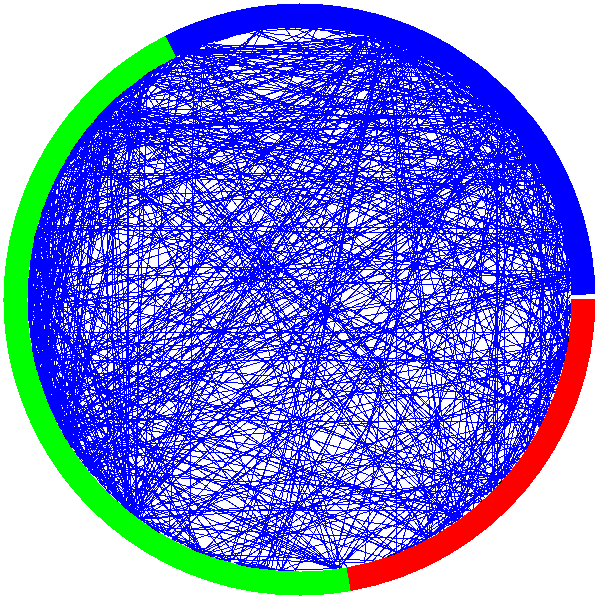}~
    &
    \includegraphics[width=0.09\columnwidth]{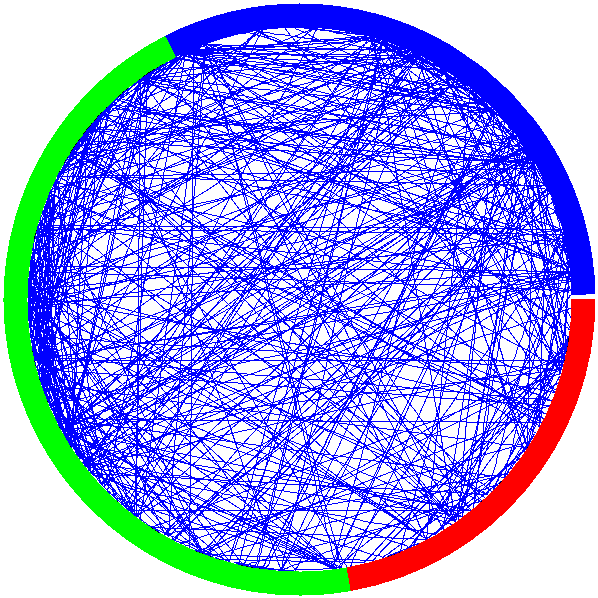}~
    &
    \includegraphics[width=0.09\columnwidth]{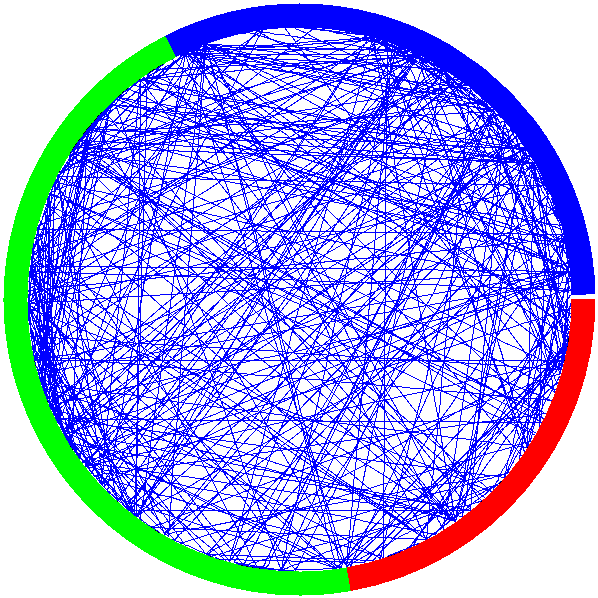}~
    &
    \includegraphics[width=0.09\columnwidth]{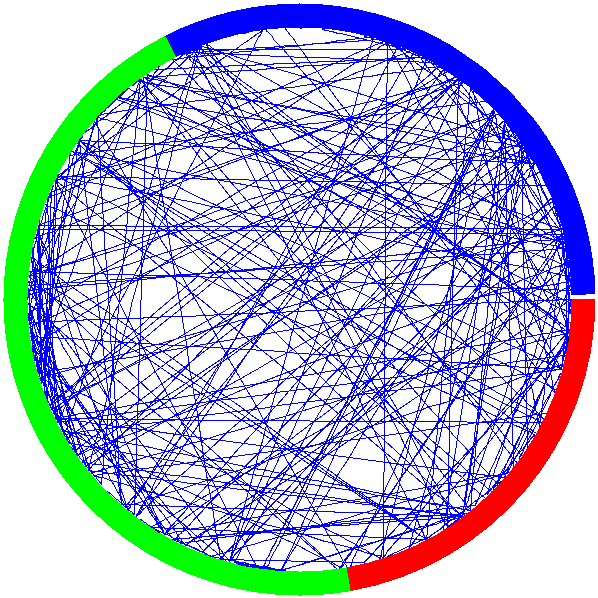}~
    &
    \includegraphics[width=0.09\columnwidth]{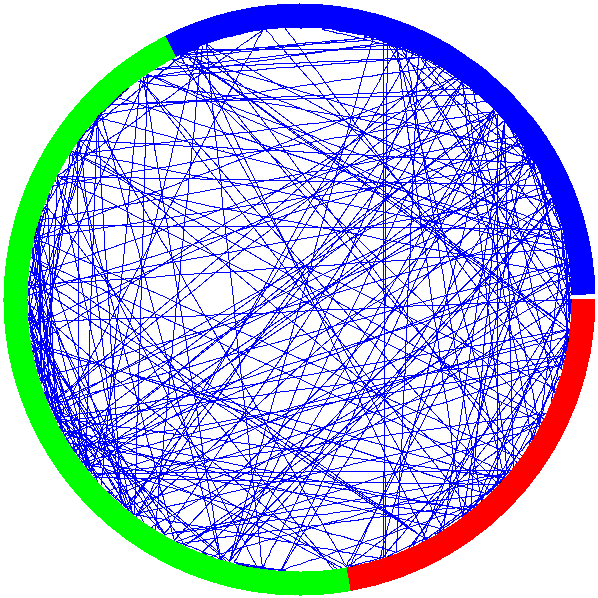}~
    &
    \includegraphics[width=0.09\columnwidth]{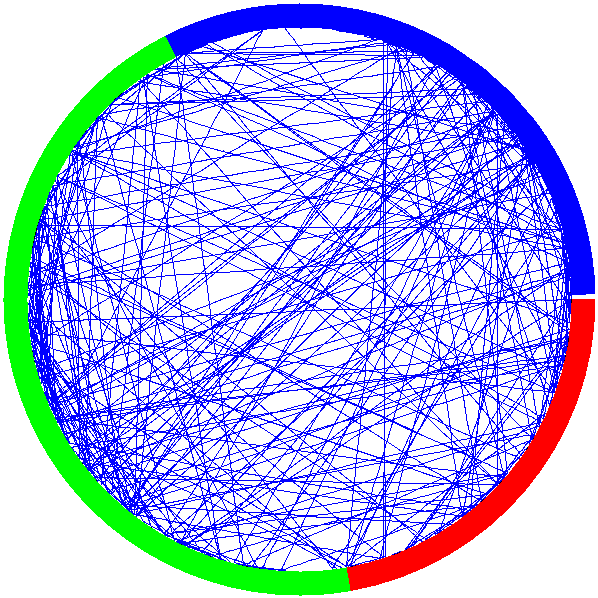}~
    \\
    \includegraphics[width=0.09\columnwidth]{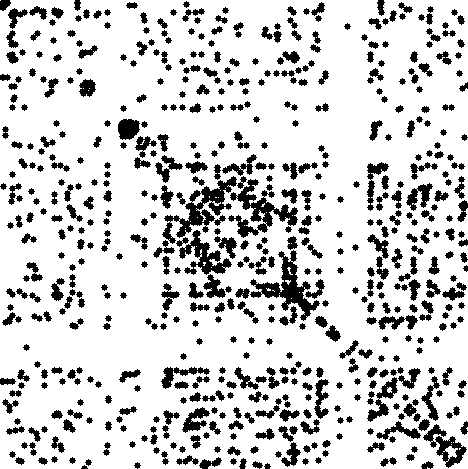}1
    &
    \includegraphics[width=0.09\columnwidth]{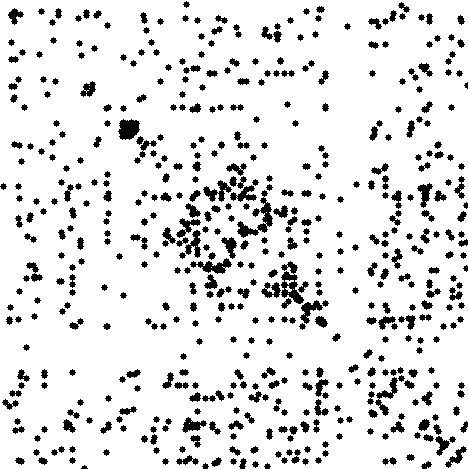}2
    &
    \includegraphics[width=0.09\columnwidth]{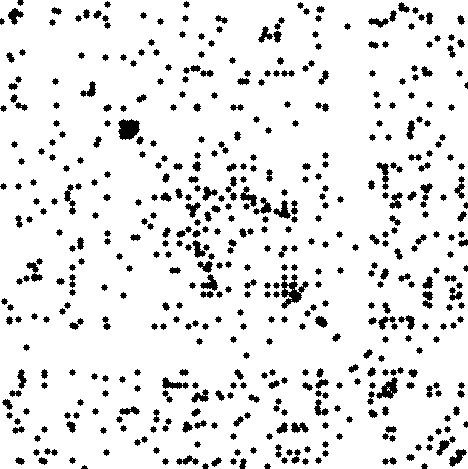}3
    &
    \includegraphics[width=0.09\columnwidth]{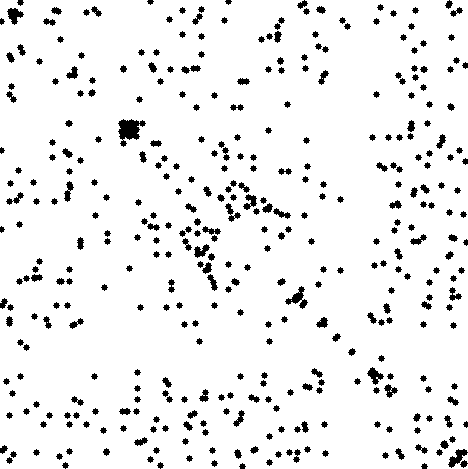}4
    &
    \includegraphics[width=0.09\columnwidth]{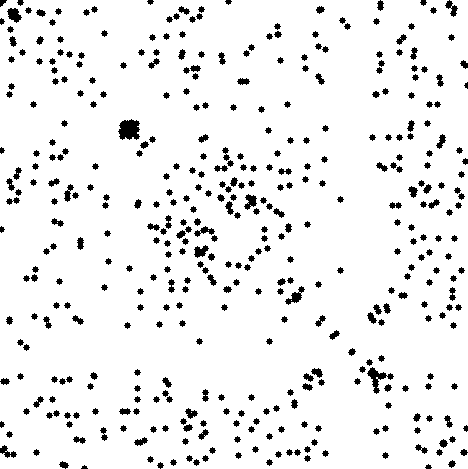}5
    &
    \includegraphics[width=0.09\columnwidth]{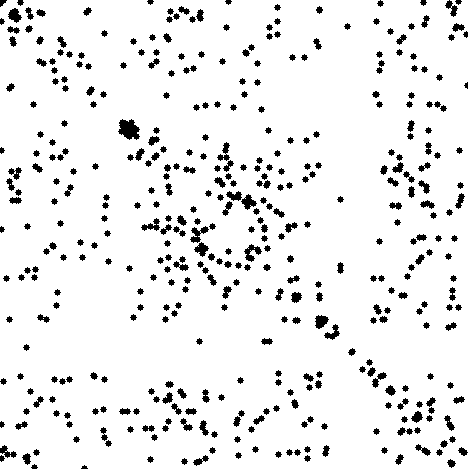}6
    &
    \\
    \cline{1-6}
    \includegraphics[width=0.09\columnwidth]{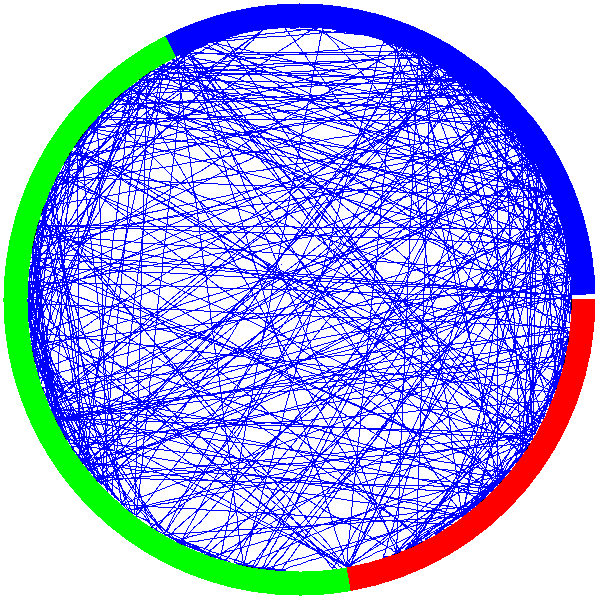}~
    &
    \includegraphics[width=0.09\columnwidth]{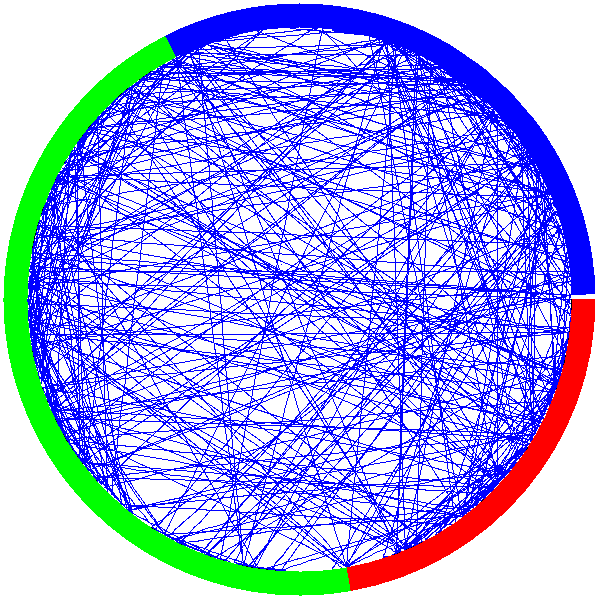}~
    &
    \includegraphics[width=0.09\columnwidth]{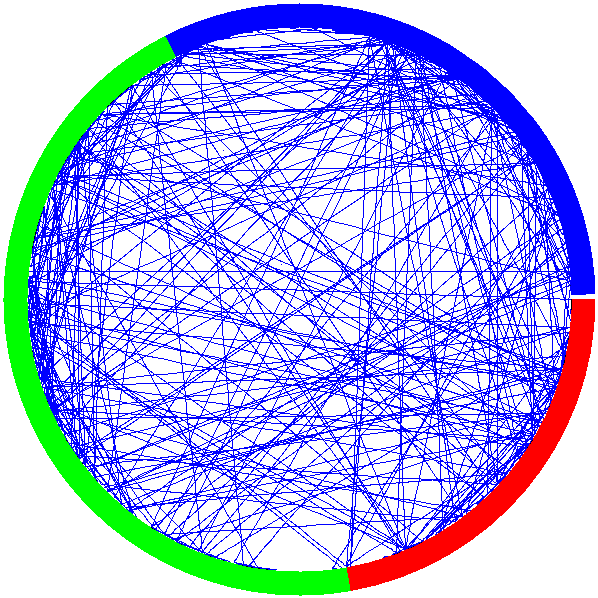}~
    &
    \includegraphics[width=0.09\columnwidth]{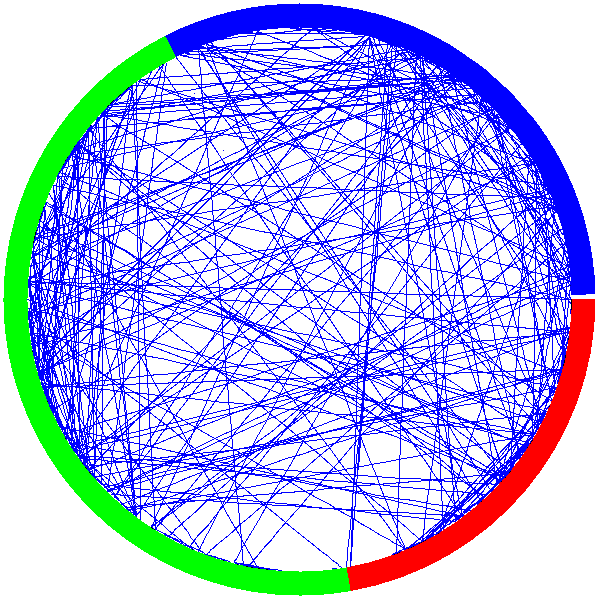}~
    &
    \includegraphics[width=0.09\columnwidth]{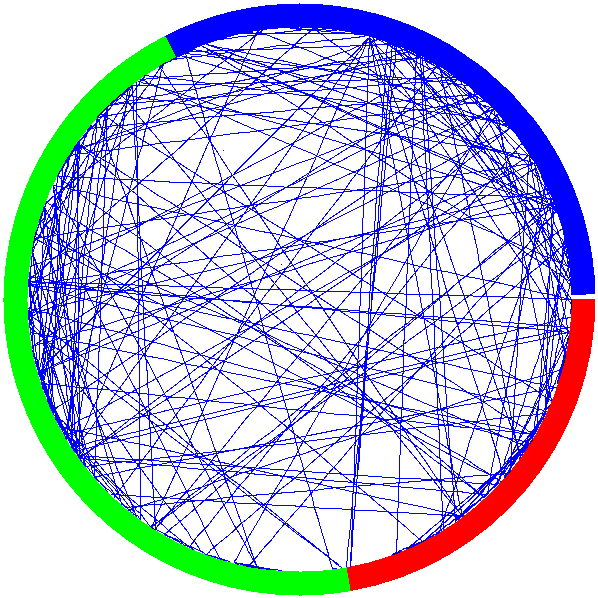}~
    &
    \includegraphics[width=0.09\columnwidth]{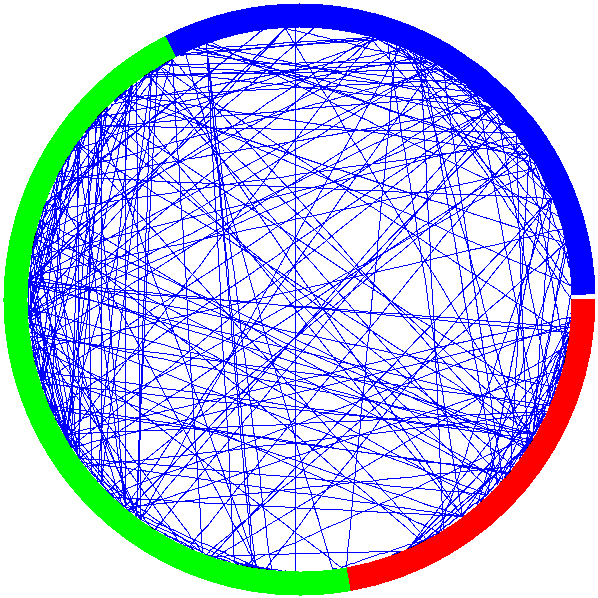}~
    \\
    \includegraphics[width=0.09\columnwidth]{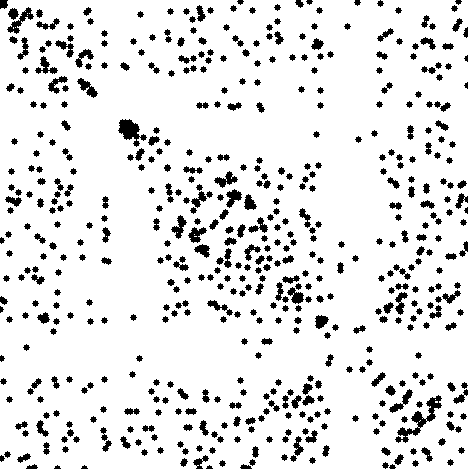}7
    &
    \includegraphics[width=0.09\columnwidth]{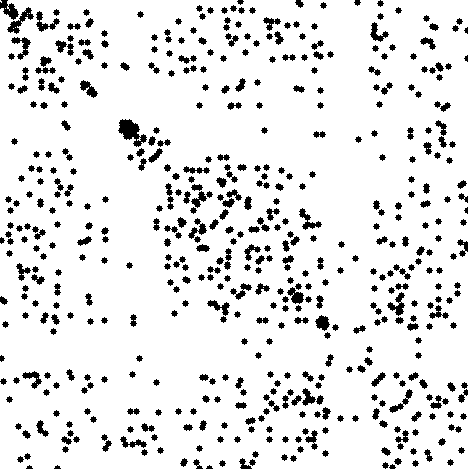}8
    &
    \includegraphics[width=0.09\columnwidth]{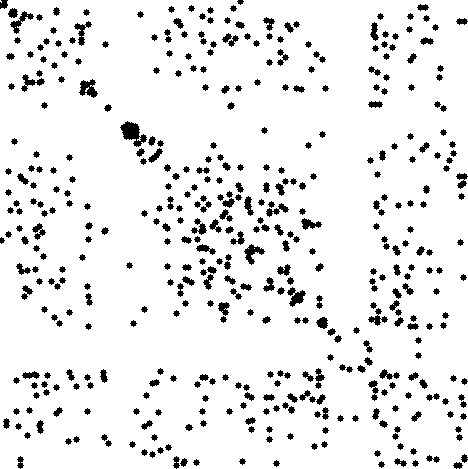}9
    &
    \includegraphics[width=0.09\columnwidth]{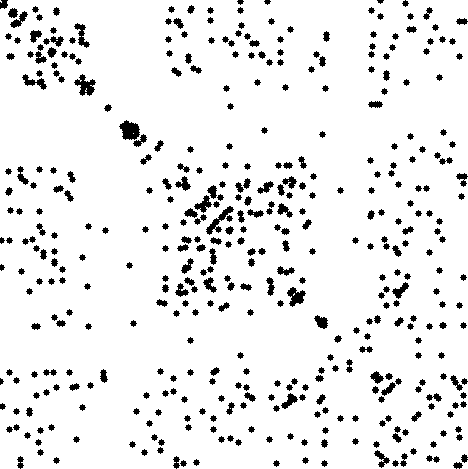}10
    &
    \includegraphics[width=0.09\columnwidth]{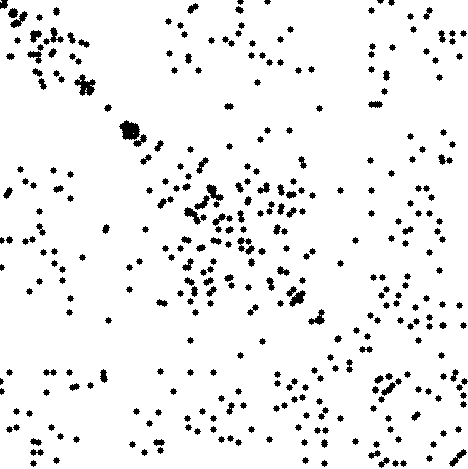}11
    &
    \includegraphics[width=0.09\columnwidth]{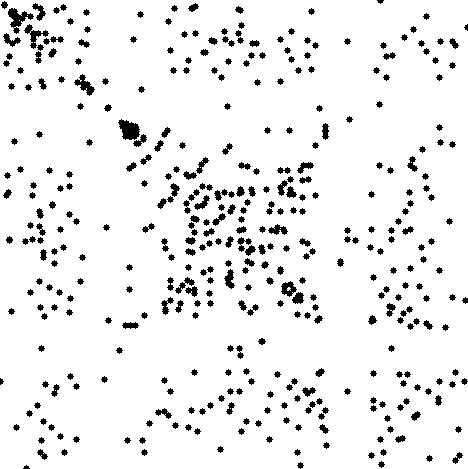}12
    &
    Visualization for the network at time point 4
    \\
    \hline
    &&&&&&
    \multirow{5}{*}{\includegraphics[width=0.28\columnwidth]{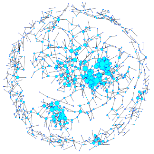}}
    \\
    \includegraphics[width=0.09\columnwidth]{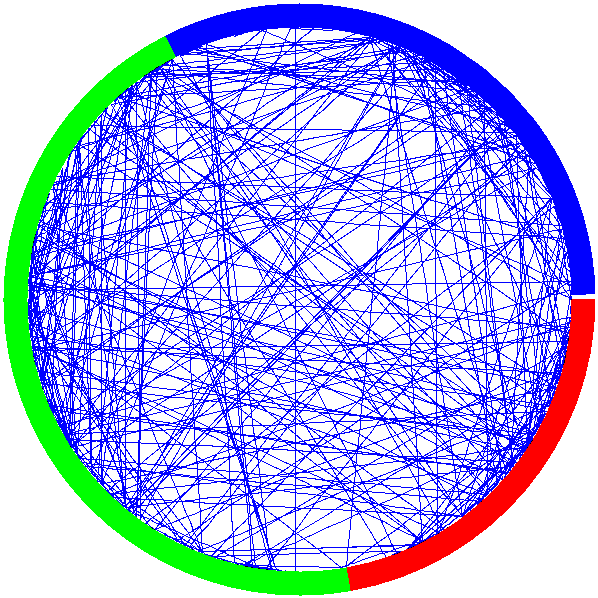}~
    &
    \includegraphics[width=0.09\columnwidth]{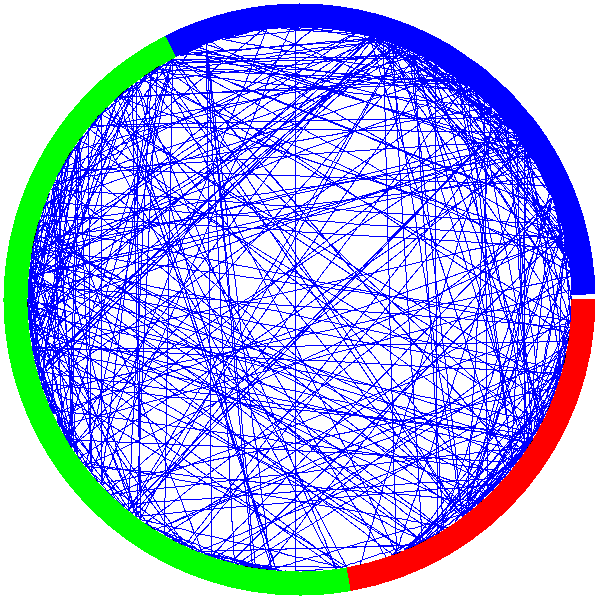}~
    &
    \includegraphics[width=0.09\columnwidth]{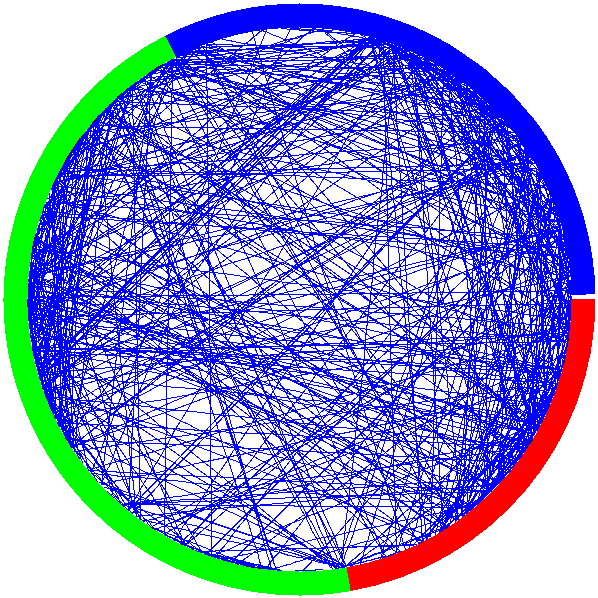}~
    &
    \includegraphics[width=0.09\columnwidth]{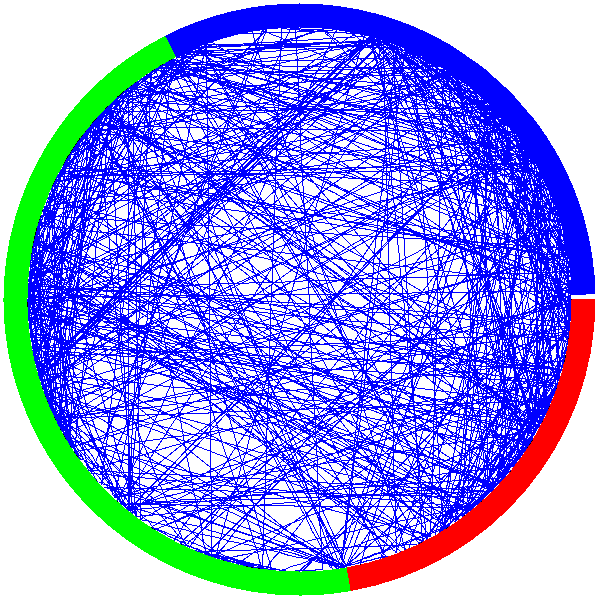}~
    &
    \includegraphics[width=0.09\columnwidth]{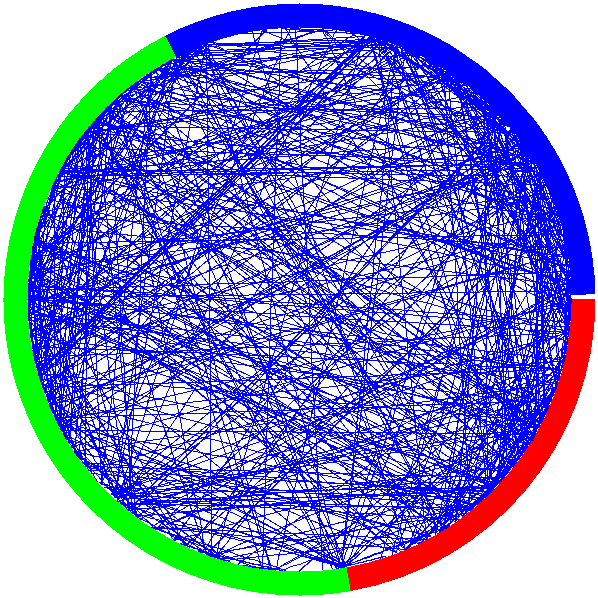}~
    &
    \includegraphics[width=0.09\columnwidth]{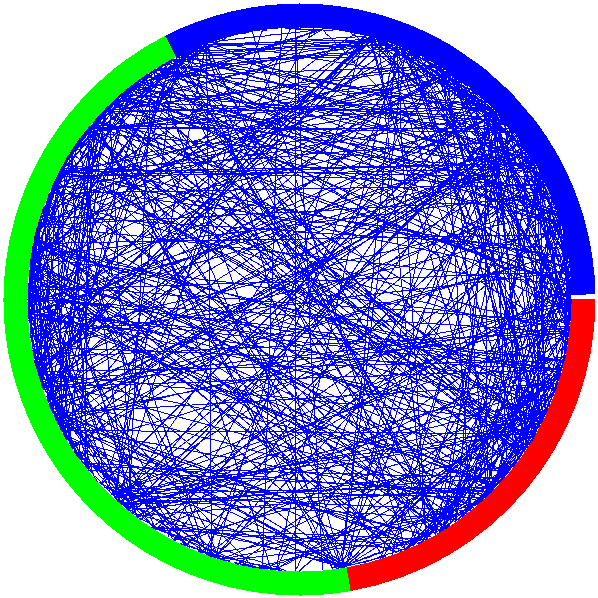}~
    \\
    \includegraphics[width=0.09\columnwidth]{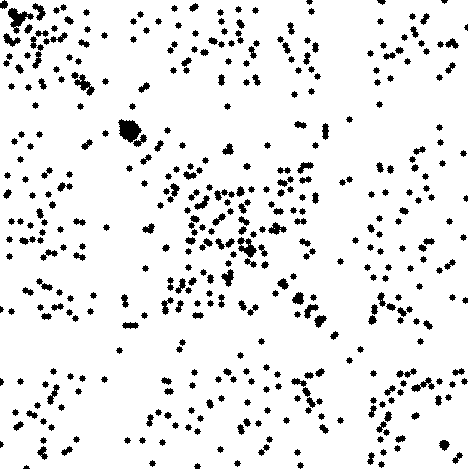}13
    &
    \includegraphics[width=0.09\columnwidth]{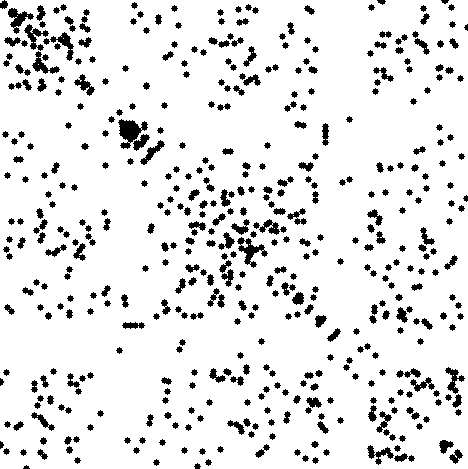}14
    &
    \includegraphics[width=0.09\columnwidth]{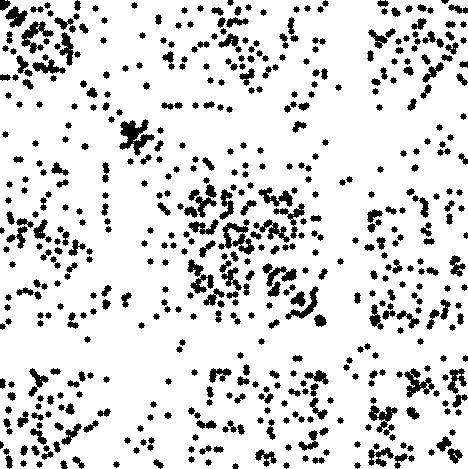}15
    &
    \includegraphics[width=0.09\columnwidth]{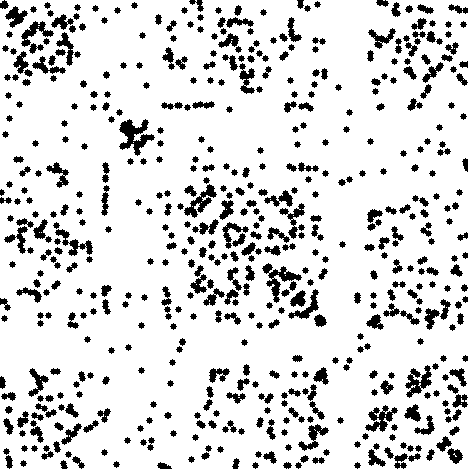}16
    &
    \includegraphics[width=0.09\columnwidth]{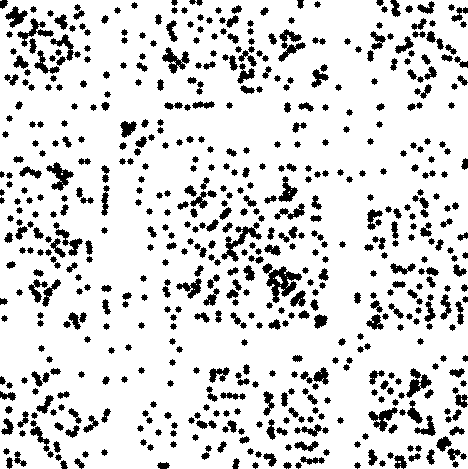}17
    &
    \includegraphics[width=0.09\columnwidth]{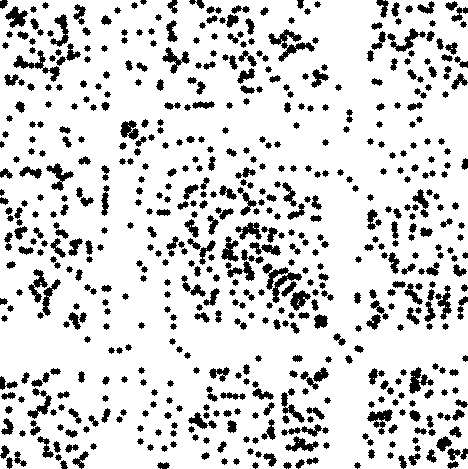}18
    &
    \\
    \cline{1-6}
    \includegraphics[width=0.09\columnwidth]{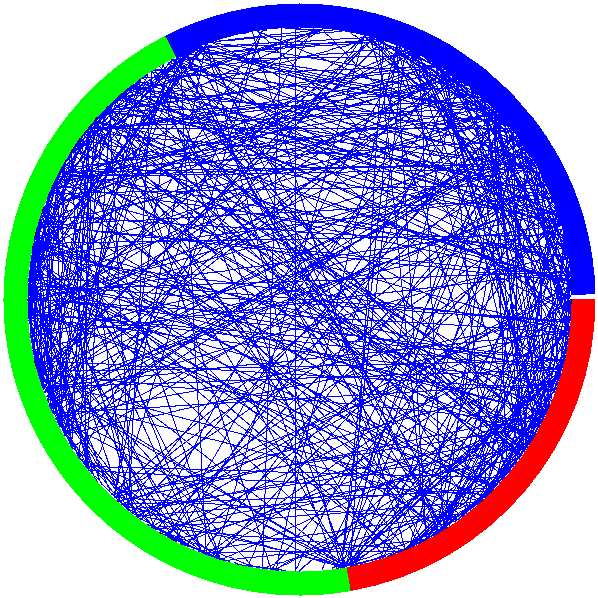}~
    &
    \includegraphics[width=0.09\columnwidth]{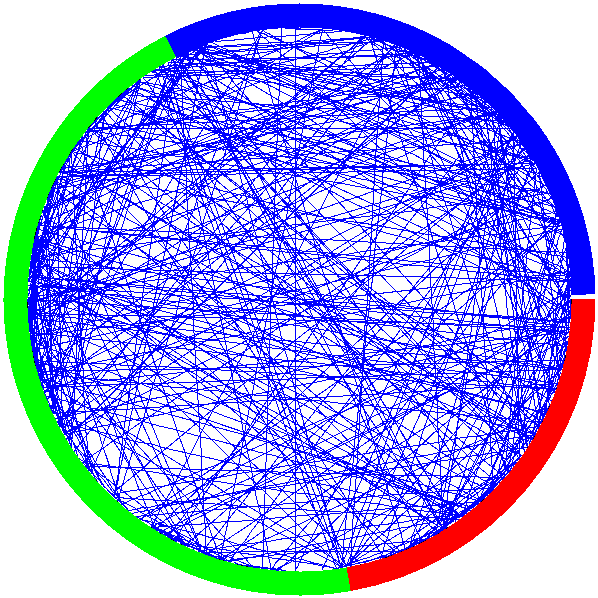}~
    &
    \includegraphics[width=0.09\columnwidth]{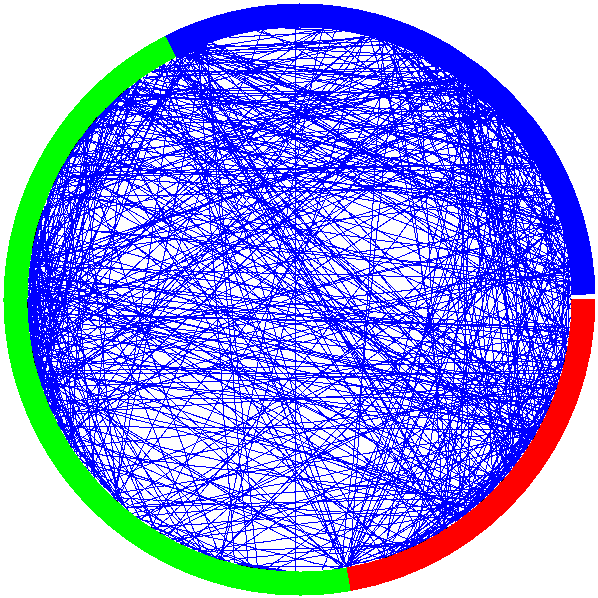}~
    &
    \includegraphics[width=0.09\columnwidth]{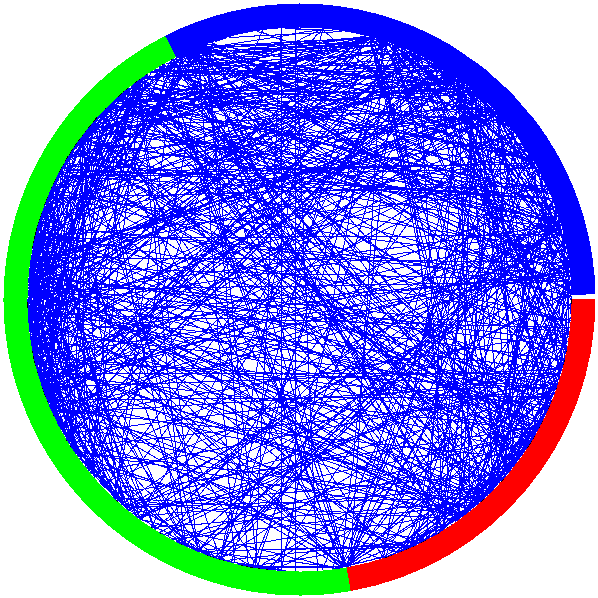}~
    &
    \includegraphics[width=0.09\columnwidth]{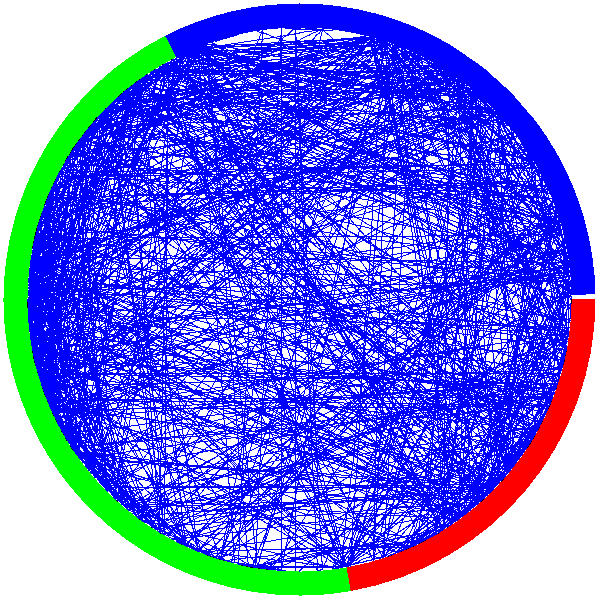}~
    &
    \includegraphics[width=0.08\columnwidth]{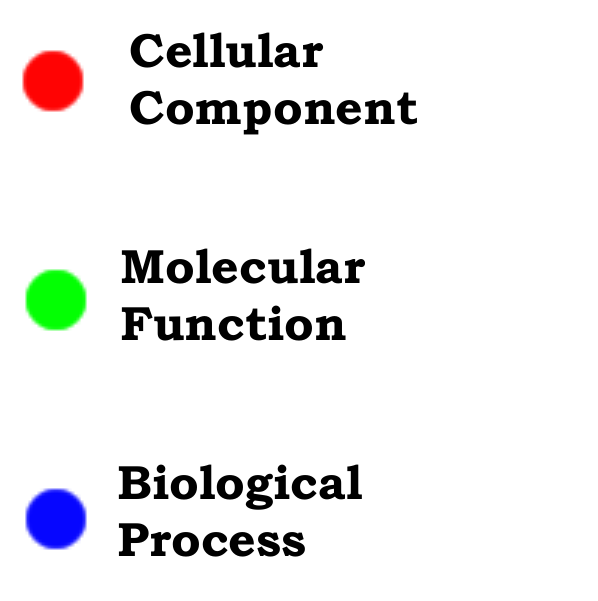}
    &
    \\
    \includegraphics[width=0.09\columnwidth]{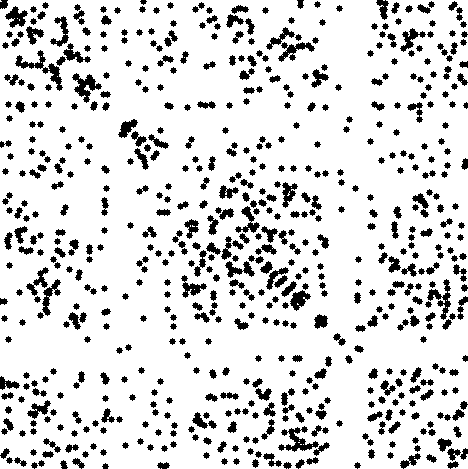}19
    &
    \includegraphics[width=0.09\columnwidth]{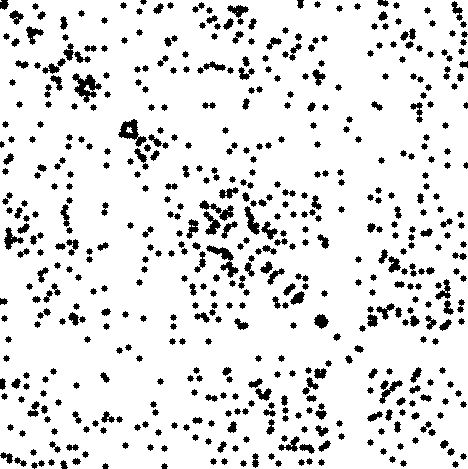}20
    &
    \includegraphics[width=0.09\columnwidth]{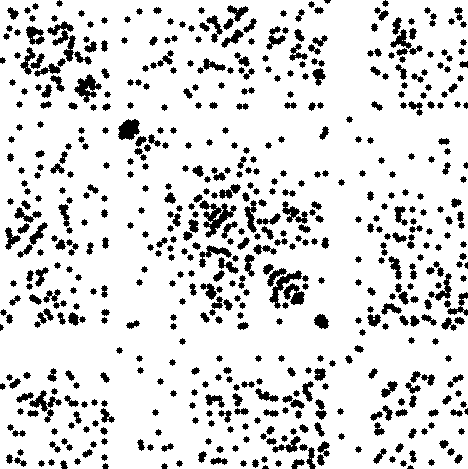}21
    &
    \includegraphics[width=0.09\columnwidth]{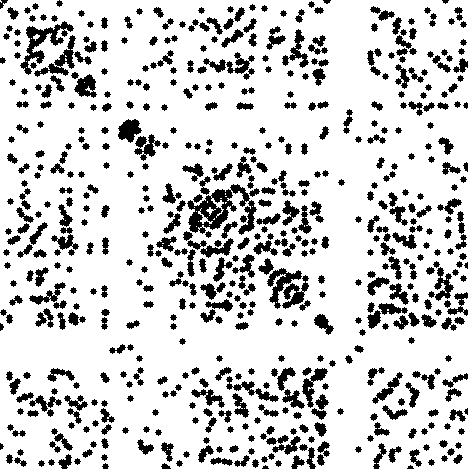}22
    &
    \includegraphics[width=0.09\columnwidth]{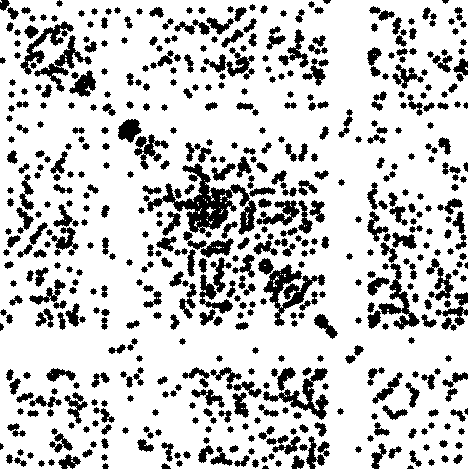}23
    &
    &
    Visualization for the network at time point 23
    \\
    \hline
    \hline
    \end{tabular}
    \caption{The 23 snapshots of the dynamic networks are
    visualized using a circular layout and a scatter plot of the network adjacency
    matrix. For both layouts, genes are ordered according to
    their top level functions (either related to cellular
    component, molecular function or biological process).
    Furthermore, two snapshots are further visualized using a
    spring layout algorithm and displayed in the last
    column of the figure.}
    \label{fg:dynamic}
\end{sidewaysfigure}

\begin{figure}
  \includegraphics[width=0.29\columnwidth]{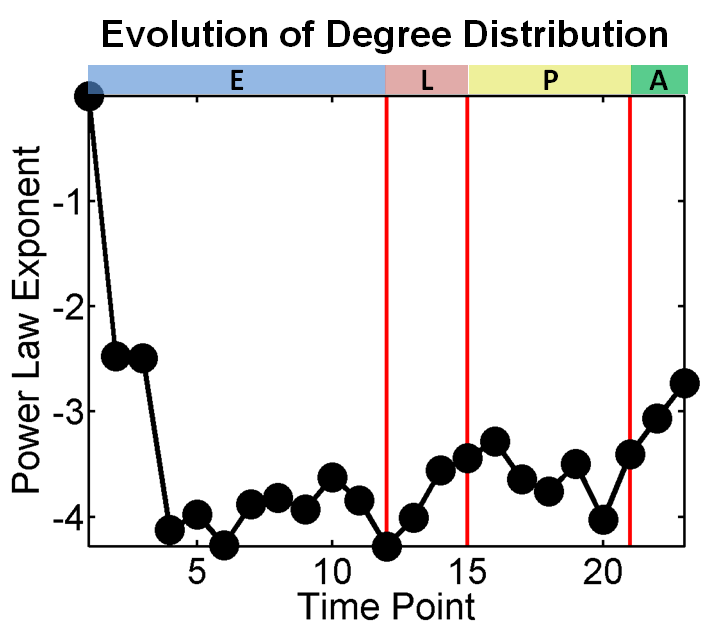}(a)
  \includegraphics[width=0.29\columnwidth]{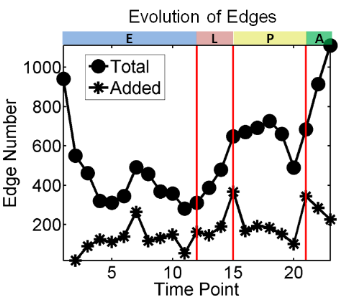}(b)
  \includegraphics[width=0.29\columnwidth]{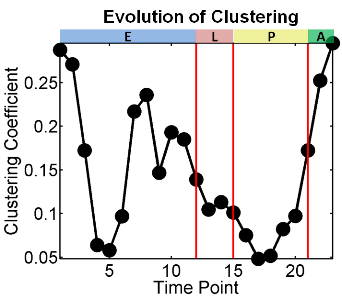}(c)\\
  \caption{The evolution of the distribution on
  the node degree is summarized using its power law exponent in (a).
  The evolution of the number
  of edge is plotted in (b). The networks
  change in size over time with new edges appearing and
  old ones disappearing. The number of newly added
  edges in each time point is also plotted in (b). Plotted in (c)
  are the clustering coefficients for each snapshot of the
  temporal networks.}
  \label{fg:networkstats}
\end{figure}

Tesla can reconstruct time-varying dynamic gene networks at a
temporal resolution from one network per every time point to one
network per epoch of arbitrary number of contiguous time points.
From the 64-step long Drosophila melanogaster life cycle microarray
time course, we reconstructed 23 dynamic gene networks, one per 3
time points, spanning the embryonic (time point 1--11), larval (time
point 12--14), pupal (time point 15--20) and adulthood stage (time
point 21--23) during the life cycle of \textit{Drosophila
melanogaster} (Fig.~\ref{fg:dynamic}) . The dynamic networks appear
to rewire over time in response to the developmental requirement of
the organism. For instance, in middle of embryonic stage (time point
4), most genes selectively interact with other genes which results
in a sparse network consisting mainly of paths. In contrast, at late
adulthood stage (time point 23), genes are more active and each gene
interacts with many other genes which leads to visible clusters of
gene interactions. The global patterns of the evolution of the gene
interactions in this network series are summarized in
Fig~\ref{fg:networkstats}. The network statistics include a summary
of the degree distribution, network size in term of the number of
edges, and the clustering coefficient of the networks, which are
computed for each snapshot of the temporally rewiring networks. The
degree distribution provides information on the average number and
the extent that one gene interacts with other genes; the network
size characterizes the total number of active gene interactions in
each time point; and the clustering coefficient quantifies the
degree of coherence inside potential functional modules. The change
of these three indices reflects the temporally rewiring nature of
the underlying gene networks.

All three indices display a wave pattern over time with these
indices peaking at the start of the embryonic stage and near the end
of adulthood stage. However, between these two stages, the
evolutions of these indices can be different. More specifically, the
degree distribution follows a similar path of evolution to that of
the network size: there are three large network changes in the
middle of embryonic stage (near time point 5), at the beginning of
the larval stage (time point 11) and at the beginning of the
adulthood stage (time point 20). In contrast, clustering
coefficients evolve in a different pattern: in the middle of the
life cycle there is only one peak which occurs near the end of the
embryonic stage. This asynchronous evolution of the clustering
coefficient from the degree distribution and network size suggests
that increased gene activity is not necessarily related to
functional coherence inside gene modules.

\begin{figure}
  \includegraphics[width=0.96\columnwidth]{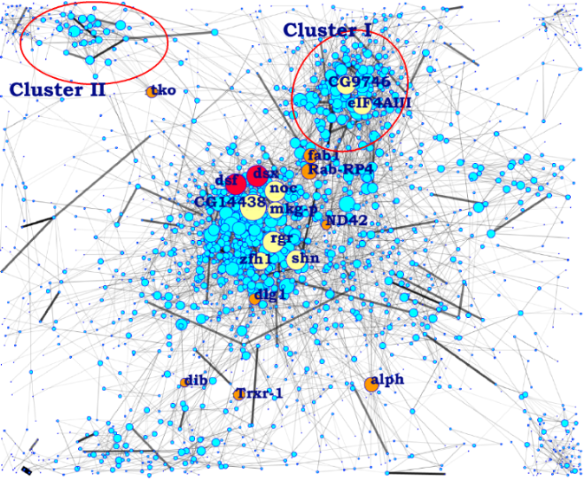}
  \includegraphics[width=0.96\columnwidth]{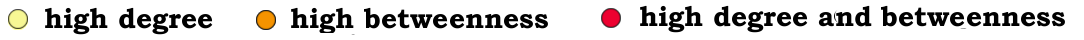}(a)\\
  \includegraphics[width=0.47\columnwidth]{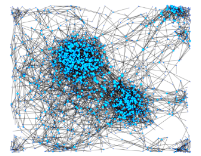}(b)
  \includegraphics[width=0.47\columnwidth]{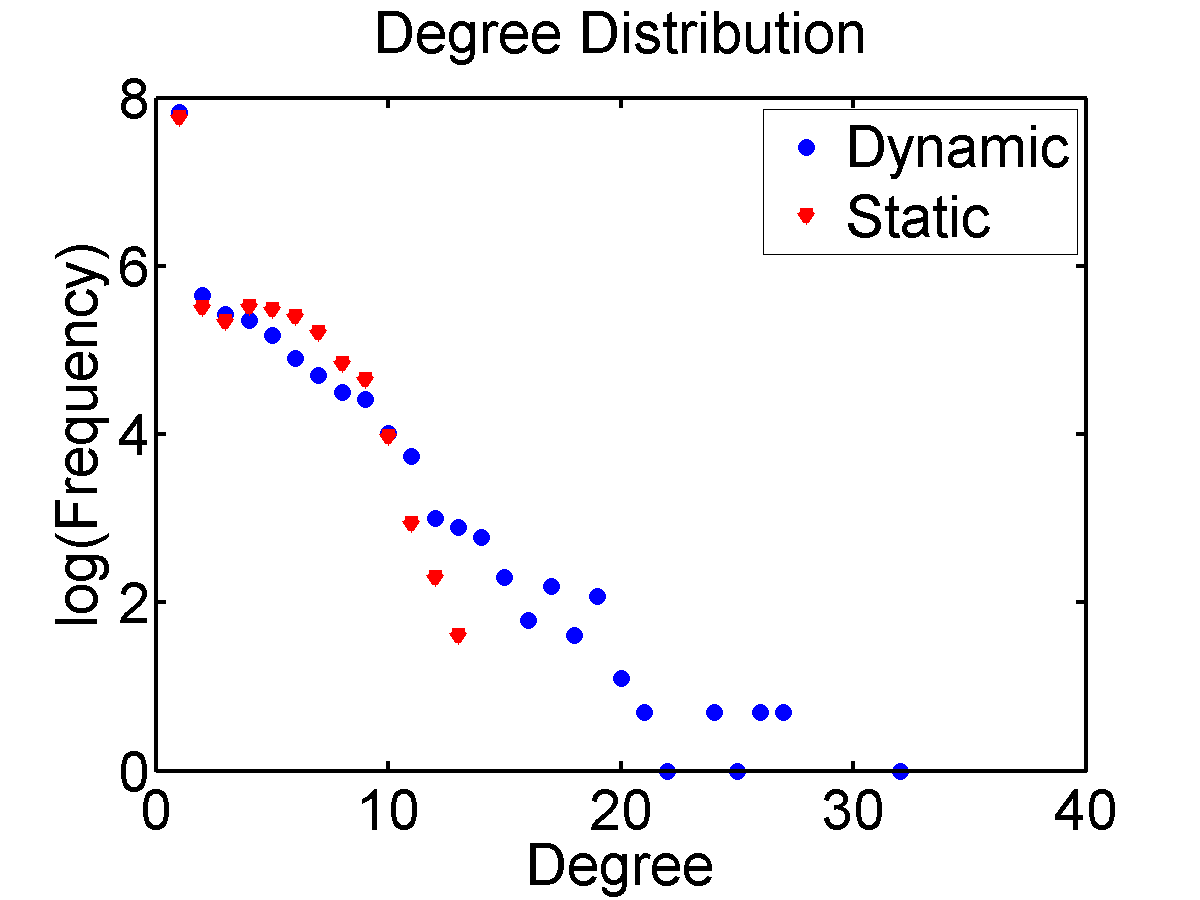}(c)\\
  \caption{The summary visualization in (a) is created by adding the 23-epoch dynamic networks
  together; the width of an edge is proportional to the number of
  times it occurs, and the size of a node is
  proportional to its degree. (b) A static network inferred by treating
  all microarray data as independently and identically distributed
  samples. (c) The degree distribution of the static network is different from that of the summary
  graph of the dynamic networks.
  }
  \label{fg:summary}
\end{figure}

To provide a summary of the 23-epoch dynamic networks, gene
interactions recovered at different time points are combined into
a single network (Fig.~\ref{fg:summary}(a)). The resulting summary
network consists of 4509 distinctive interactions. Overall the
summary network reveals a center of two giant clusters of tightly
interacting genes, with several small loose clusters remotely
connected to the center. The two giant clusters are each built
around genes with high degree centrality. These two clusters and
other small clusters are integrated as a single network via a set
of genes with high betweenness centrality.

The top giant cluster is centered around two high degree nodes,
protein coding gene eIF4AIII and CG9746. Both genes are related to
molecular functions such as ATP binding. The selective interaction
of these genes with other molecules modulates the biological
functions of these molecules. In the bottom giant cluster, high
degree nodes are more abundant compared to the top cluster. These
high degree nodes usually have more than one functions. For
instance, gene dsf and dsx are involved in DNA binding and
modulate their transcription, which in turn regulates the sex
related biological processes. Gene noc, CG14438, mkg-p are
involved in both ion binding and intracellular component, while
gene zfh1 and shn play important roles in the developmental
process of the organism.

The connections between the clusters are very often channeled
through a set of genes with high betweenness centrality. Note
these genes do not necessarily have a high degree. They are
important because they provide the relays for many biological
pathways. For instance, gene fab1 is involved in various molecular
functions such as ATP, protein and ion binding, while at the same
time it is also involved in biological processes such as
intracellular signaling cascade. Gene dlg1 participate in various
functions such as protein binding, cell polarity determination and
cellular component. Another gene tko participates in functions
related to ribosome from both molecular, biological and cellular
component point of views.

For comparison, a single static network of comparable size
($\sim$4500 gene interactions) is inferred by treating all
measurements as independent and identically distributed samples
(Fig~\ref{fg:summary}(e)). The static network shares some common
features with the summary graph of the dynamic networks. For
instance, both networks reveal two giant clusters of interacting
genes. However, the static network provides no temporal information
on the onset and duration of each gene interaction. Furthermore, the
static network and the summary graph of the dynamic networks follow
very different degree distributions. The degree distribution of the
summary graph of the dynamic networks has a much heavier tail than
that of the static network (Fig~\ref{fg:summary}(f)). In other
words, the distribution for the dynamic networks resembles more to a
scale free network while that for the static network resembles more
to a random graph.

\subsection*{Hub Genes in the Dynamic Networks}
\begin{sidewaysfigure}
    \begin{tabular}{cccc}
        \multirow{2}{*}{\includegraphics[width=0.21\columnwidth]{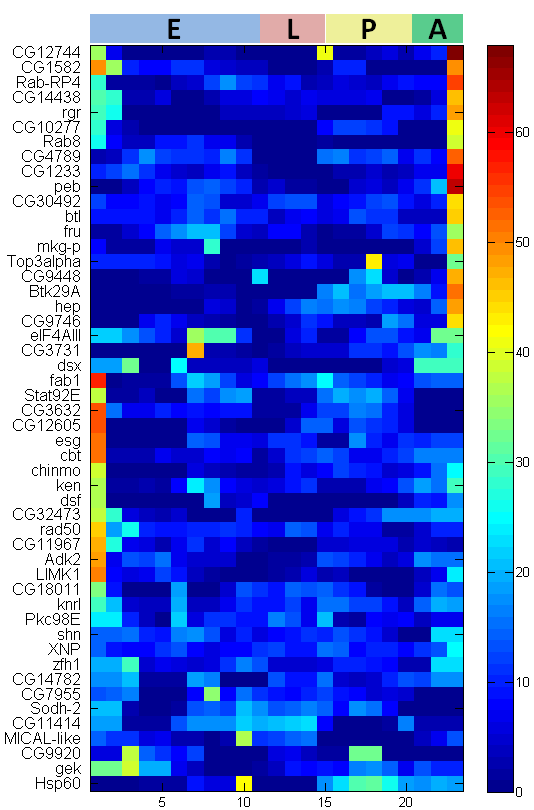}(a)}
        &
        \multirow{2}{*}{\includegraphics[width=0.22\columnwidth]{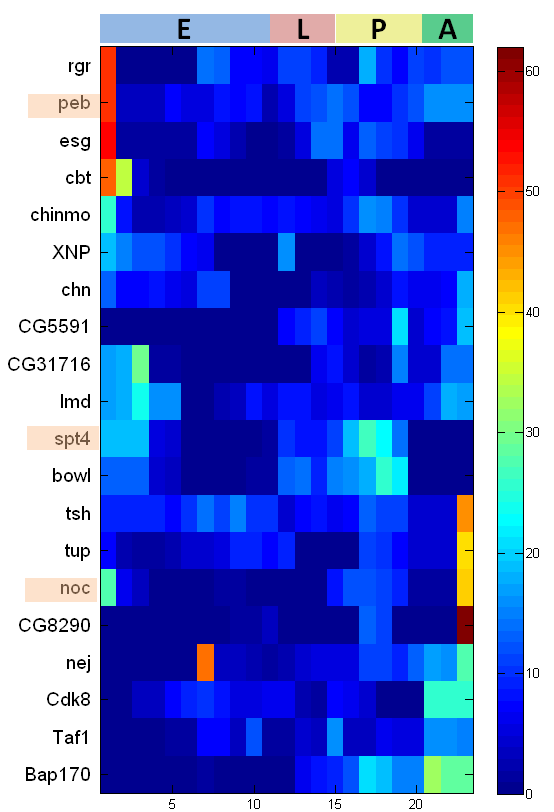}(b)}
        &&
        \\
        &&
        \includegraphics[width=0.22\columnwidth]{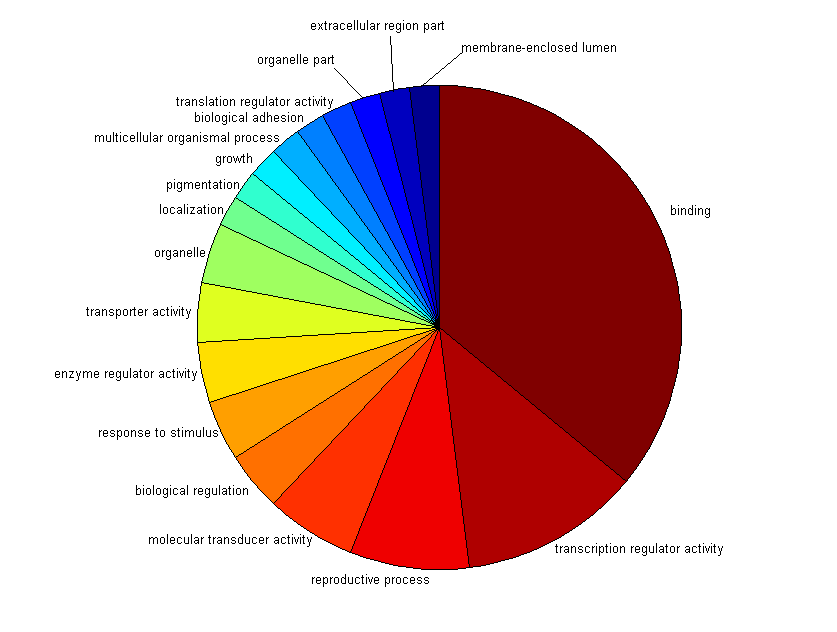}(c) &
        \includegraphics[width=0.22\columnwidth]{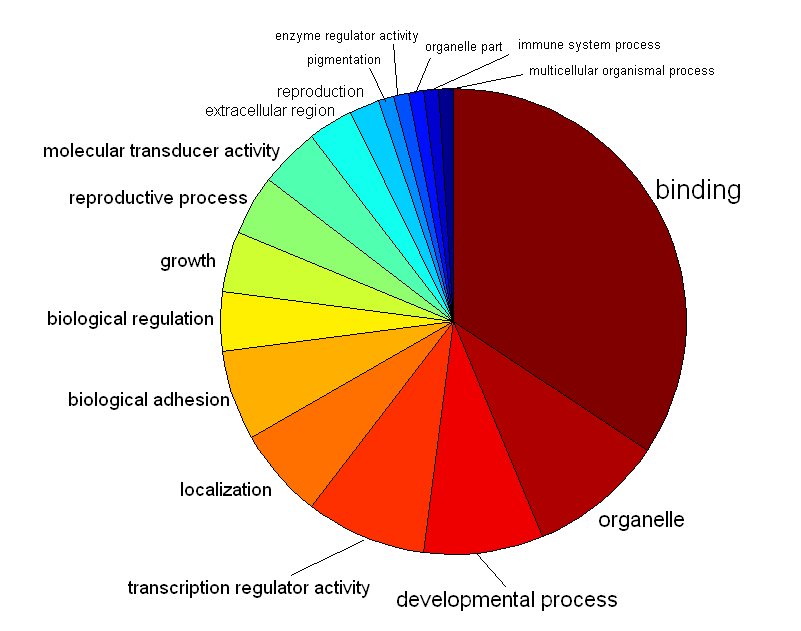}(d)
        \\
        &&
        \includegraphics[width=0.22\columnwidth]{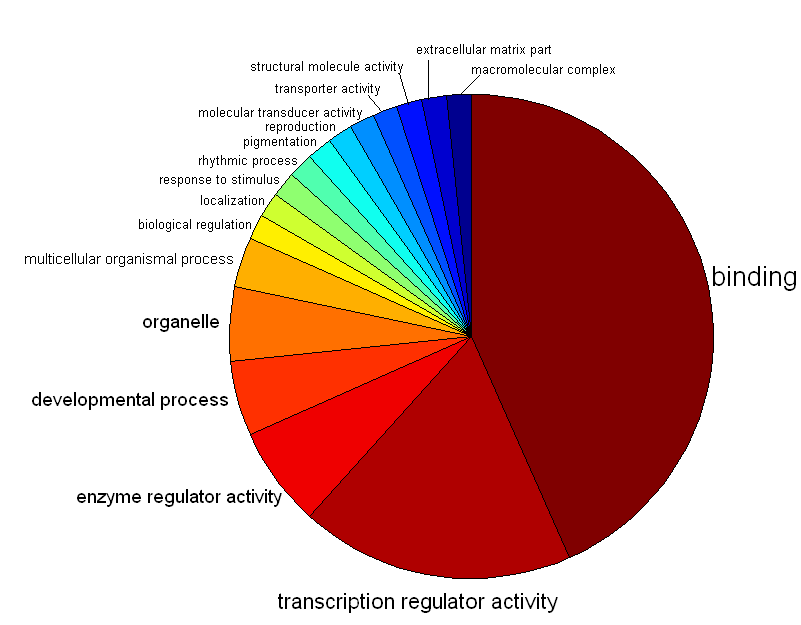}(e) &
        \includegraphics[width=0.22\columnwidth]{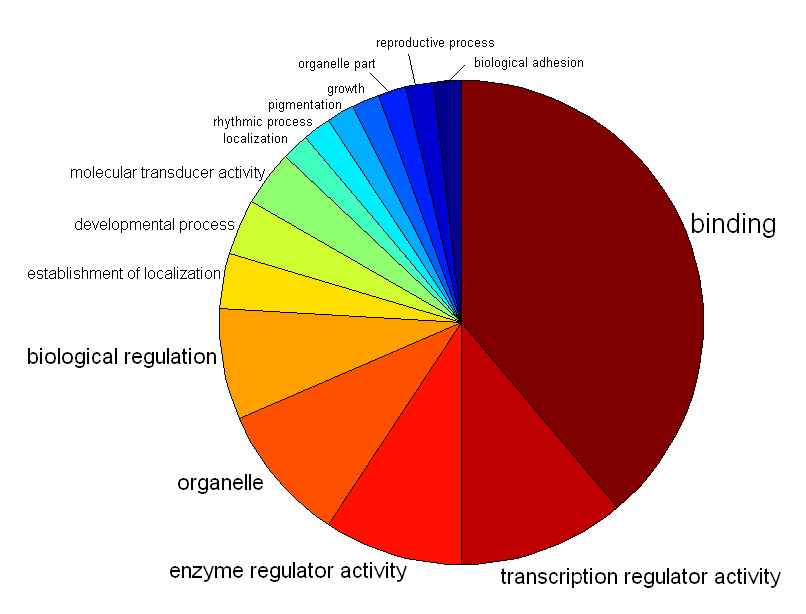}(f)
        \\
    \end{tabular}
    \caption{(a) The top 50 hubs in the summary graph of the dynamic networks are tracked over time
    in terms of their degrees. Each row represents one hub and each column
    represents a time point with the color code corresponds to the degree of the
    hub. (b) shows the same information as (a) but only for the top 20
    transcriptional factors with largest degree. (c) shows the functional decomposition of the top 50 hubs.
    (d)(e)(f) show examples of functional decomposition of the genes regulated by the
    transcriptional factor hub. These examples correspond to the highlighted gene in (b)
    and their gene names are peb, spt4 and noc respectively.
    }
    \label{fig:hubs}
\end{sidewaysfigure}

The hub genes correspond to the high degree nodes in summary
network. They represent the most influential elements of a network
and tend to be essential for the developmental process of the
organism. The top 50 hubs are identified from the summary graph of
the dynamic networks in Fig.~\ref{fg:summary}. These hubs are
tracked over time in terms of their degrees and this evolution are
visualized as colormap in Fig.~\ref{fig:hubs}(a).

To further understand the role played by these hubs, histogram
analysis are performed on these 50 hubs
in term of 43 ontological functions. The functional decomposition
of these hubs are shown in Fig.~\ref{fig:hubs}(c).
The majority of these hubs are related to functions such as binding and
transcriptional regulation activity. This is in fact an expected
outcome as transcription factors (TF) are thought to target a large
number of genes and modulate their expression.

To further understand
the functional spectrum of the genes targeted by the high degree transcriptional factors,
the top 20 transcriptional factor hubs are also tracked over time in terms of their degrees
and the evolution is illustrated in Fig.
\ref{fig:hubs}(b). The degrees of the transcriptional factors peak at different stages
which means they differentially trigger target genes based on the
biological requirements of developmental process. In
Fig. \ref{fig:hubs}(d)(e)(f), functional decomposition is performed
on the target genes regulated by three example transcriptional factor hubs. For instance, peb, the
protein ejaculatory bulb, interacts with extracellular region
genes and genes involved in structural molecular activity.
Another example is spt4 which triggers many binding genes.
This is consistent with its functional role in chromatin binding and
zinc ion binding.

\subsection*{Dynamic Clustering of Genes}

Most gene interactions occur only at certain time during the life
cycle of \textit{Drosophila melanogaster}. Indeed, on average
there are only one eighth of the total gene interactions present
in each temporal snapshot of the dynamic networks. The clusters in
the summary graph are the result of temporal accumulation of the
dynamic networks. To illustrate this, two clusters of genes are
singled out from Fig~\ref{fg:summary}(a) for further study.
Cluster I consists of 167 genes and there are five major gene
ontology groups, ie. binding activity (34.7\%), organelle (9\%),
transporter activity (6.6\%), transducer activity (6.6\%) and
motor activity (5.4\%); cluster II consists of 90 genes and there
are three major gene ontology groups, ie. structural molecule
activity (61.1\%), organelle (8.9\%) and transporter activity
(6.7\%).

\begin{figure}
\centering
    \begin{tabular}{cc}
    \multirow{4}{*}{\includegraphics[width=0.40\columnwidth]{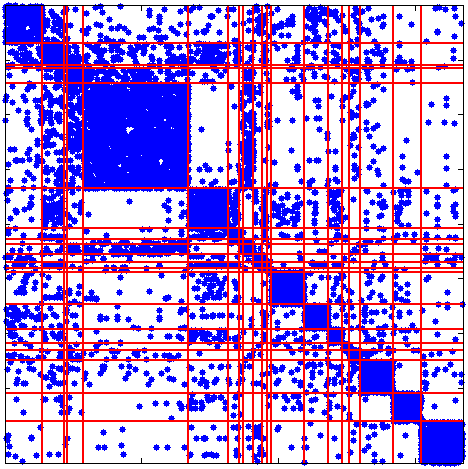}}
    &
    \\
    &
    \includegraphics[width=0.12\columnwidth]{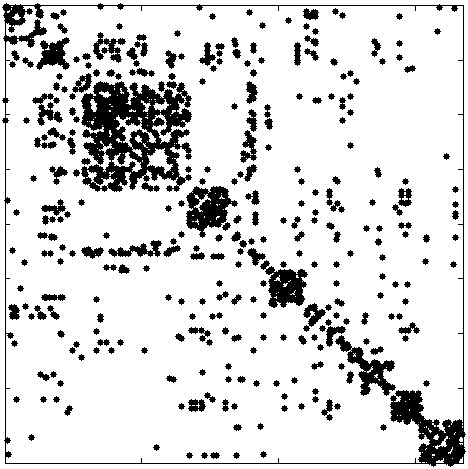}1
    \includegraphics[width=0.12\columnwidth]{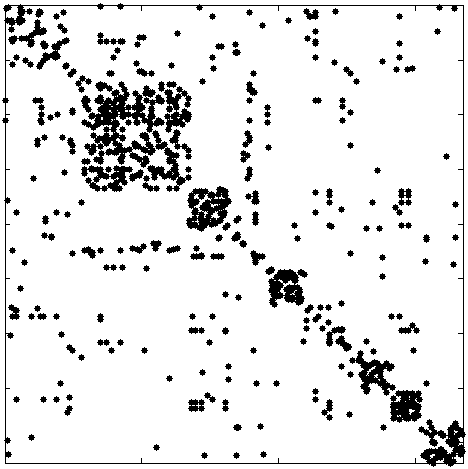}2
    \includegraphics[width=0.12\columnwidth]{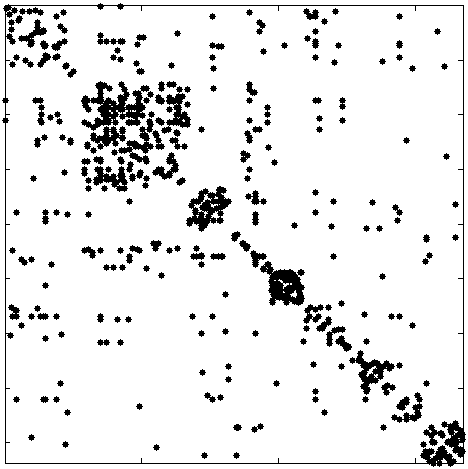}3
    \includegraphics[width=0.12\columnwidth]{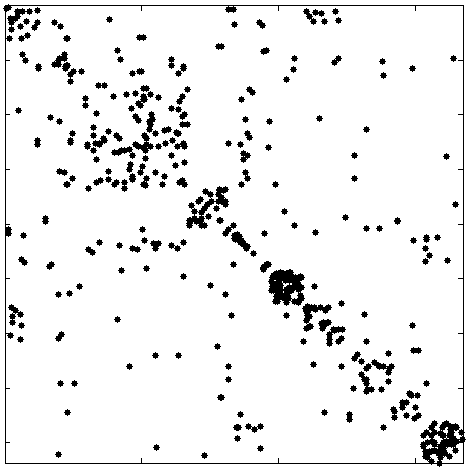}4
    \\
    &
    \includegraphics[width=0.12\columnwidth]{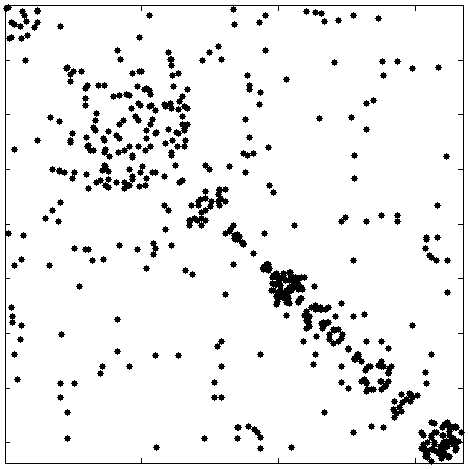}5
    \includegraphics[width=0.12\columnwidth]{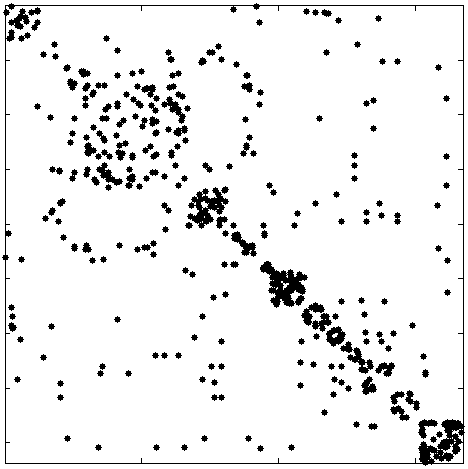}6
    \includegraphics[width=0.12\columnwidth]{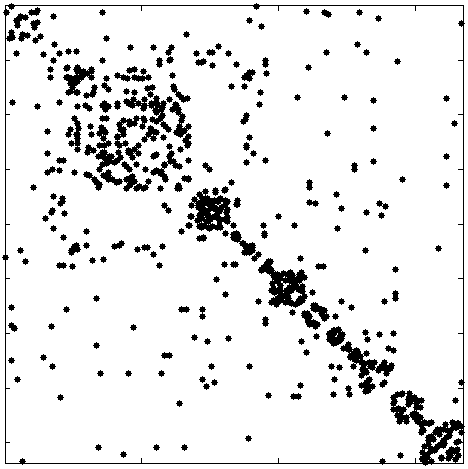}7
    \includegraphics[width=0.12\columnwidth]{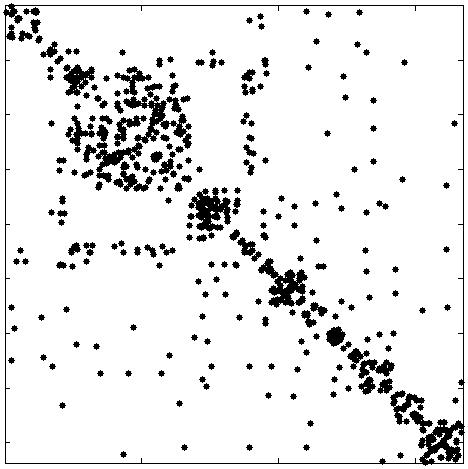}8
    \\
    &
    \includegraphics[width=0.12\columnwidth]{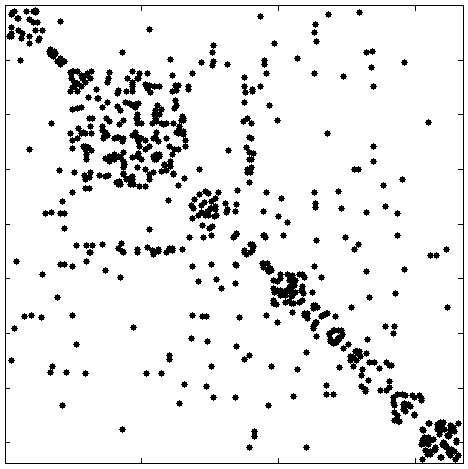}9
    \includegraphics[width=0.12\columnwidth]{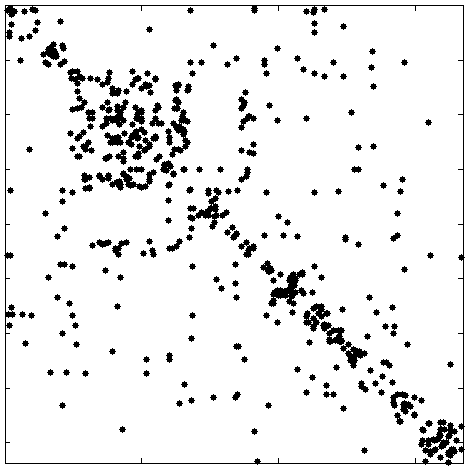}10
    \includegraphics[width=0.12\columnwidth]{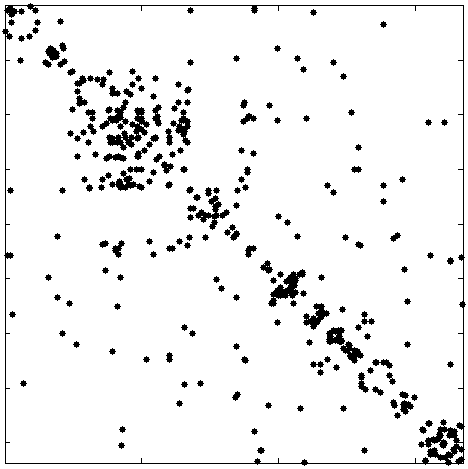}11
    \includegraphics[width=0.12\columnwidth]{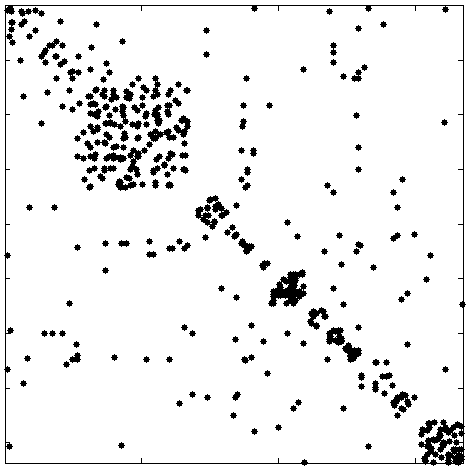}12
    \\
    &
    \includegraphics[width=0.12\columnwidth]{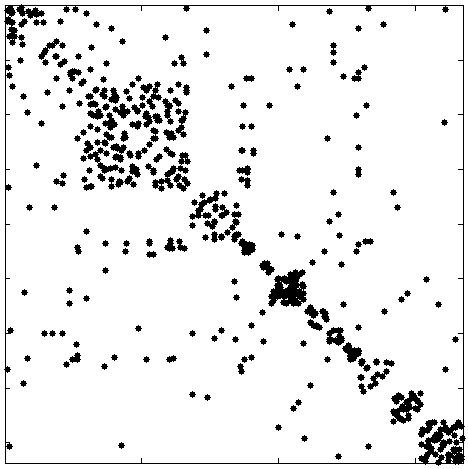}13
    \includegraphics[width=0.12\columnwidth]{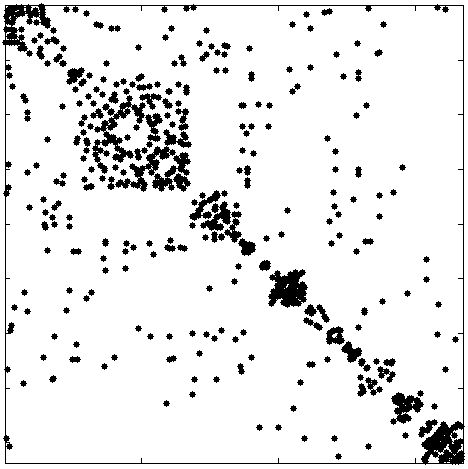}14
    \includegraphics[width=0.12\columnwidth]{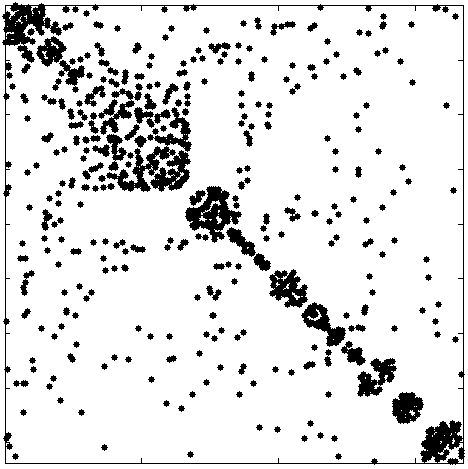}15
    \includegraphics[width=0.12\columnwidth]{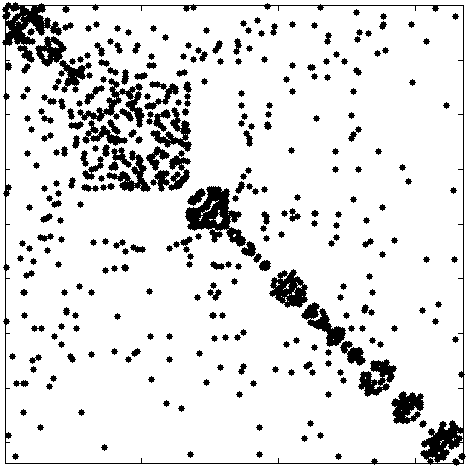}16
    \\
    \includegraphics[width=0.12\columnwidth]{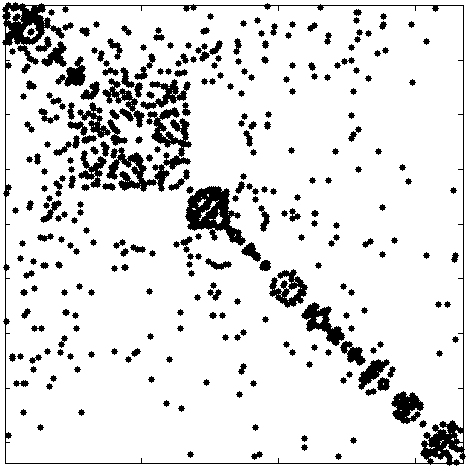}17
    \includegraphics[width=0.12\columnwidth]{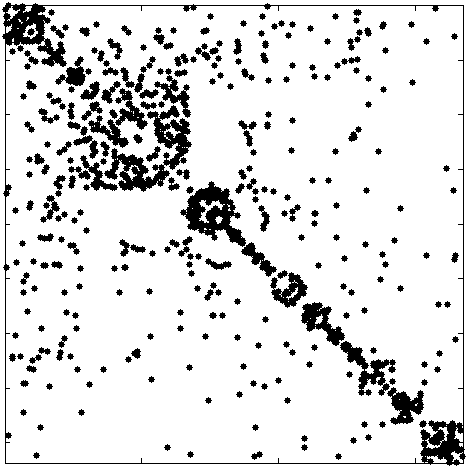}18
    \includegraphics[width=0.12\columnwidth]{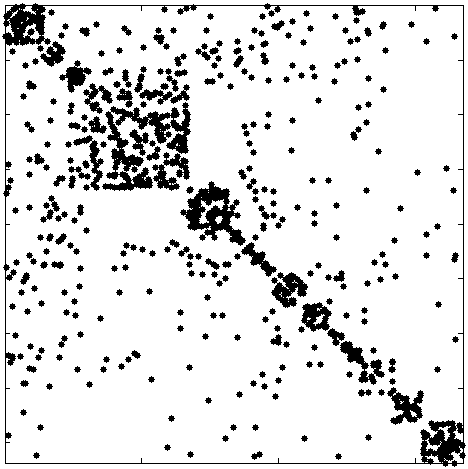}19
    &
    \includegraphics[width=0.12\columnwidth]{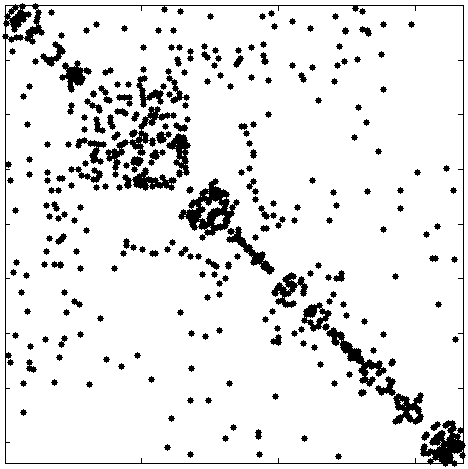}20
    \includegraphics[width=0.12\columnwidth]{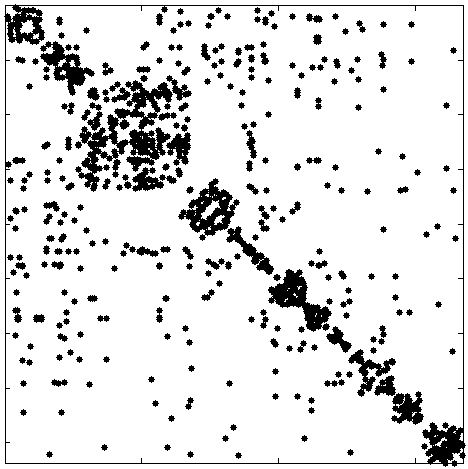}21
    \includegraphics[width=0.12\columnwidth]{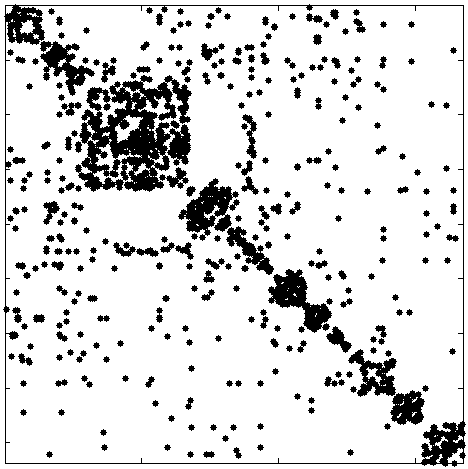}22
    \includegraphics[width=0.12\columnwidth]{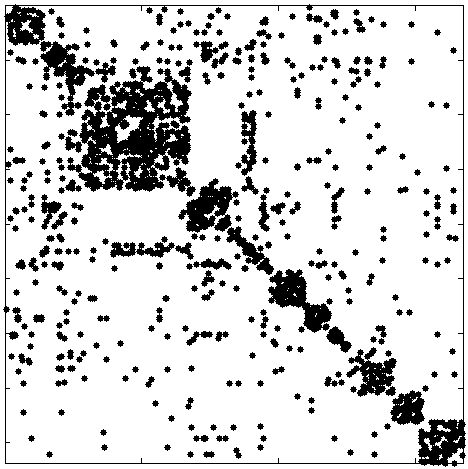}23
    \\
    \end{tabular}
    \caption{The adjacency matrix of the 20 clusters of genes
    derived from the largest connected components of
    Fig.~\ref{fg:summary} are plotted over time. Showed on the
    upper left corner is an enlarged picture of the adjacency
    matrix of the summary graph. The red lines show the boundaries
    between the clusters.}
    \label{fg:cluster}
\end{figure}

To obtain a finer functional decomposition of the genes in the
interaction networks, the largest connected component of the
summary graph is further grouped into 20 clusters. Although this
connected component consists only of about 40\% (1674) of all
genes , it contains more than 97\% (4401) of the interactions in
the summary graph. This 20 clusters vary in size, with the
smallest cluster having only 11 genes and the largest cluster
having 384 genes. The evolution of these cluster of genes is
illustrated in Fig.~\ref{fg:cluster}. It can be seen that the
connections within a cluster dissolve and reappear over time, and
different clusters wax and wane according to different schedule.

Furthermore, the functional composition of these 20 clusters are
compared against the background functional composition of the set
of all 4028 genes. For this purpose, gene ontology terms are used
as the bins for the histogram, and the number of genes belonging
to each ontology group are counted into the corresponding bins.
The functional composition of 9 out of the 20 clusters are
statistically significantly different from the background (two
sample Kolmogorov-Smirnov test at significance level 0.05). The
top 5 functional components of these significant clusters are
summarized in Table~\ref{tb:clusterfunc}.

\begin{table}
\small
  \centering
  \caption{Top 5 functional components of the 9 clusters whose functional compositions are statistically
  significantly different from the back functional composition. The second column of the table displays
  the number of genes in each cluster respectively. The functional composition of each cluster is displayed
  as the percentage of genes with that particular function. The last column of the table shows that
  $p$-value of the statistical test.}\label{tb:clusterfunc}
  \begin{tabular}{|c|c|c|c|}
    \hline
    \hline
    \# & Genes \# & Top 5 Function Components &
    $p$-value \\
    \hline
    1 & 138 & transcription regulator activity (15.9\%), developmental process (8.7\%), & $<10^{-6}$ \\
    && organelle (13.8\%), reproduction (8.0\%), localization (5.8\%) & \\
    \hline
    2 & 384 & binding (37.8\%), organelle (12.5\%), transcription regulator activity (7.8\%), & $<10^{-6}$ \\
    && enzyme regulator activity (7.6\%), biological regulation (3.6\%) & \\
    \hline
    3 & 146 & binding (29.5\%), transporter activity (11.0\%), organelle (8.9\%), & $0.05$ \\
    && motor activity (6.2\%), molecular transducer activity (5.5\%) & \\
    \hline
    4 & 120 & structural molecule activity (45.0\%), translation regulator activity (10.8\%), & $<10^{-6}$ \\
    && organelle (6.7\%), binding (5.8\%), macromolecular complex (4.2\%) & \\
    \hline
    5 & 89 & organelle (14.6\%), binding (13.5\%), multicellular organismal process (11.2\%), & $10^{-3}$ \\
    && developmental process (7.9\%), biological regulation (7.9\%) & \\
    \hline
    6 & 25 & catalytic activity (16.0\%), translation regulator activity (16.0\%), organelle (8.0\%), & 0.04 \\
    && antioxidant activity (8.0\%), cellular process (8.0\%) & \\
    \hline
    7 & 39 & extracellular region (28.2\%), biological adhesion (12.8\%), binding (7.7\%), & 0.007 \\
    && reproduction (7.7\%), transporter activity (5.1\%) & \\
    \hline
    8 & 119 & multi-organism process (16.0\%), reproduction (9.2\%), developmental process (9.2\%), & $<10^{-6}$ \\
    && molecular transducer activity (7.6\%), organelle (5.9\%) & \\
    \hline
    9 & 151 & binding (28.5\%), envelope (14.6\%), organelle (9.9\%), & $<10^{-6}$ \\
    && organelle part (9.9\%), transporter activity (5.3\%) & \\
    \hline
    \hline
  \end{tabular}
\end{table}

During the life cycle of \textit{Drosophila melanogaster}, the
developmental program of the organism may require genes related to
one function be more active in certain stage than others. To
investigate this, genes are grouped according to their ontological
functions. It is expected that the interactions between gene
ontology groups, as quantified by the number of wirings from genes
in one group to those in another group, also exhibit temporal
pattern of rewiring.

\begin{sidewaysfigure}
    \begin{tabular}{cc}
    \multirow{4}{*}{\includegraphics[width=0.40\columnwidth]{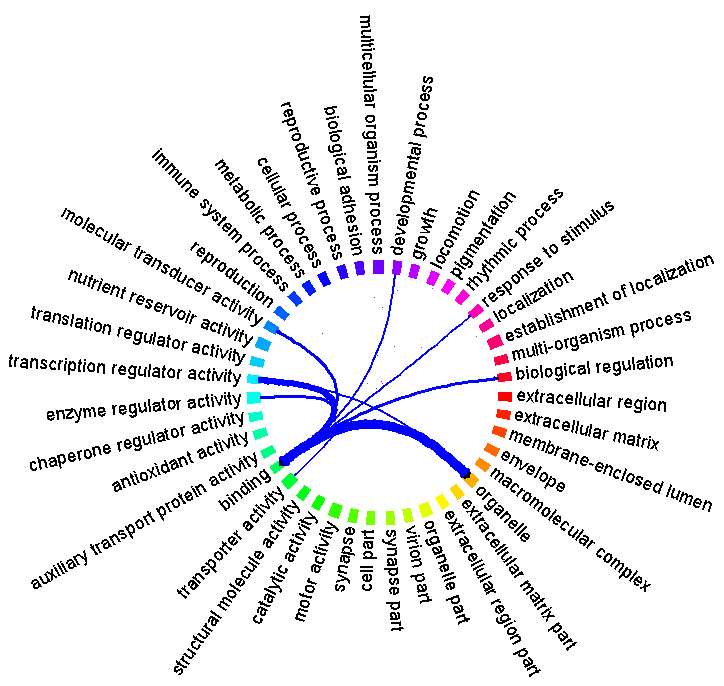}}
    &
    \\
    &
    \includegraphics[width=0.12\columnwidth]{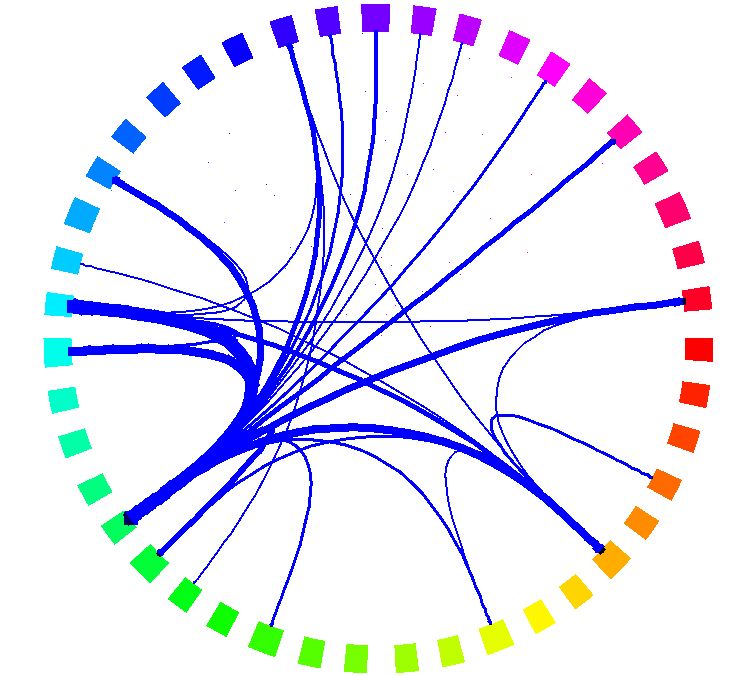}1
    \includegraphics[width=0.12\columnwidth]{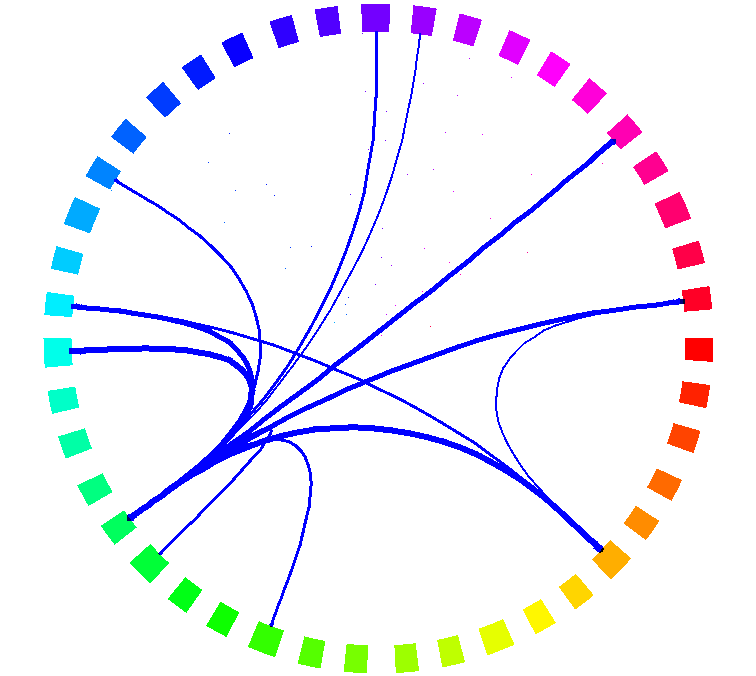}2
    \includegraphics[width=0.12\columnwidth]{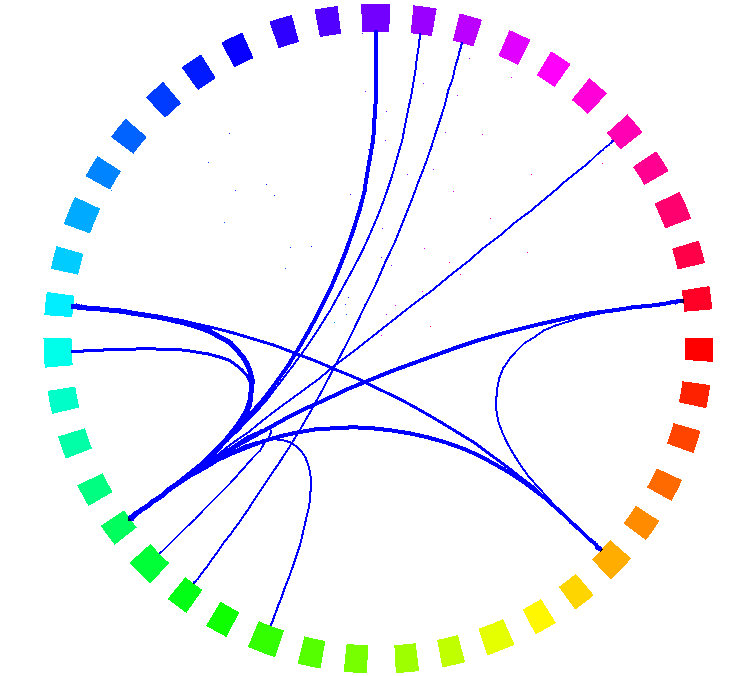}3
    \includegraphics[width=0.12\columnwidth]{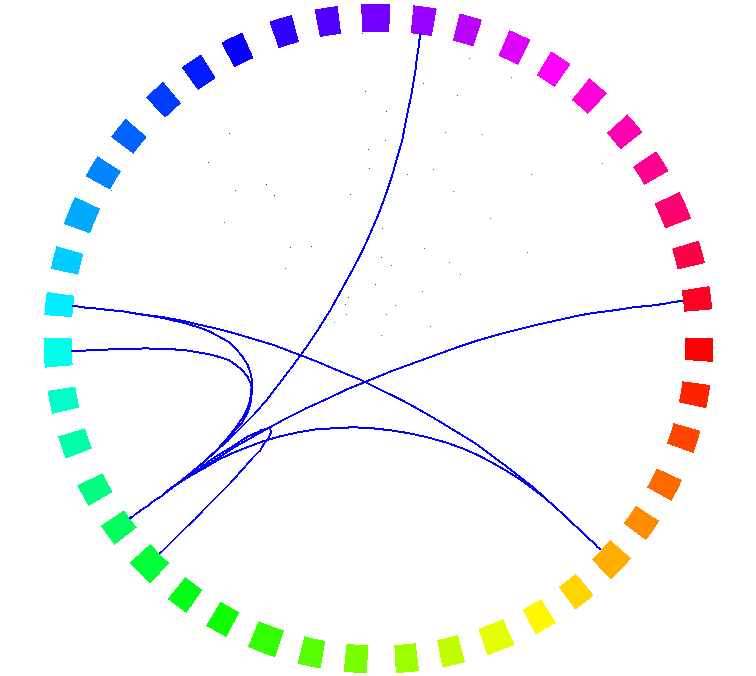}4
    \\
    &
    \includegraphics[width=0.12\columnwidth]{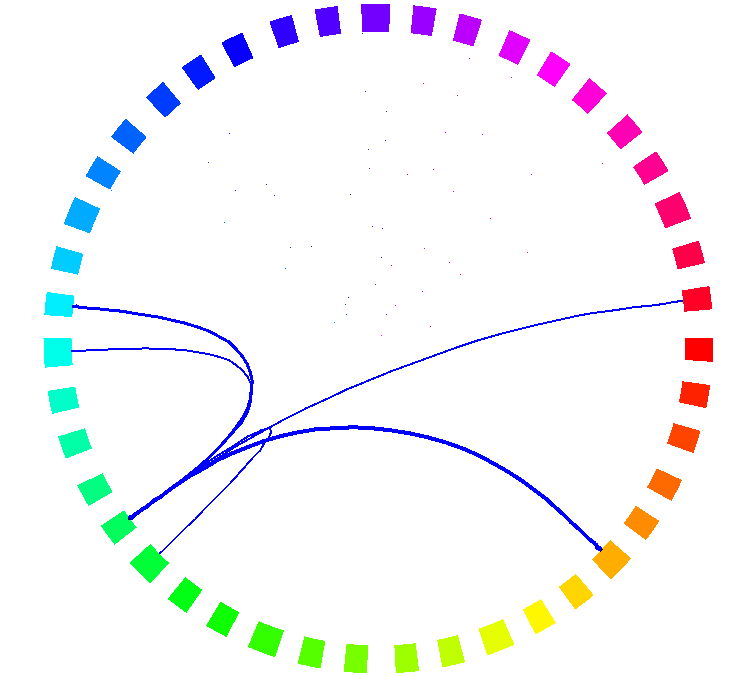}5
    \includegraphics[width=0.12\columnwidth]{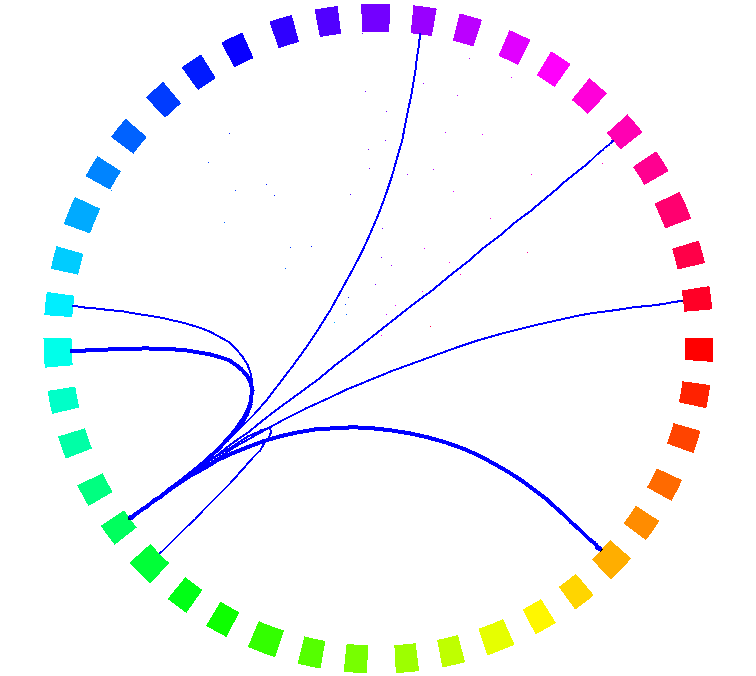}6
    \includegraphics[width=0.12\columnwidth]{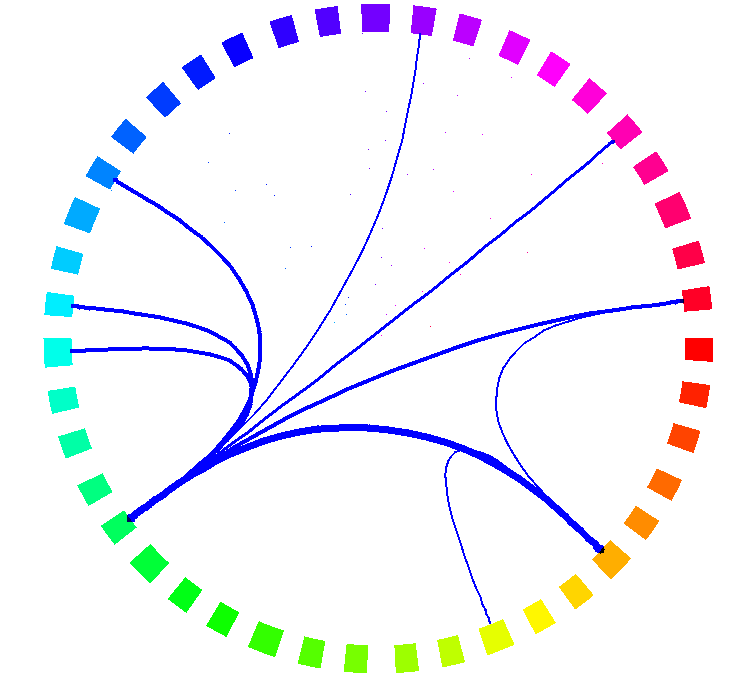}7
    \includegraphics[width=0.12\columnwidth]{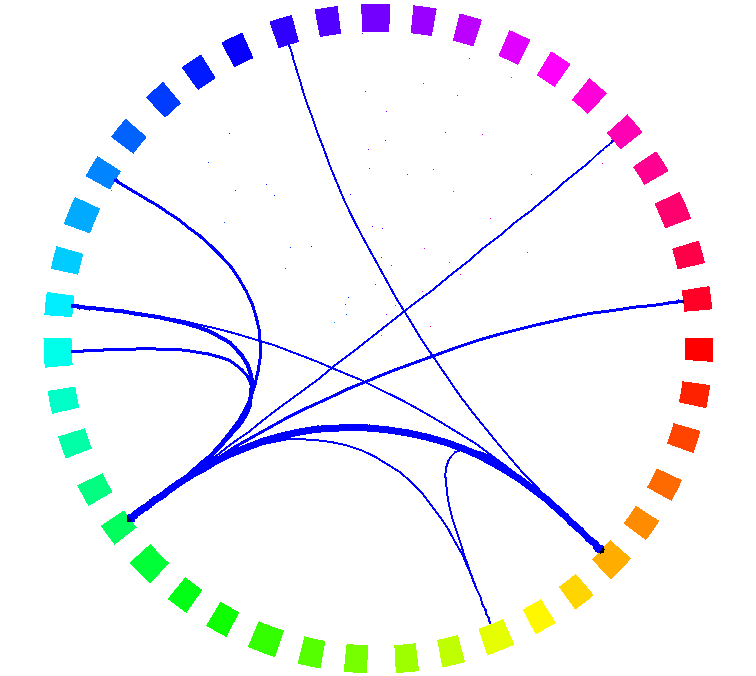}8
    \\
    &
    \includegraphics[width=0.12\columnwidth]{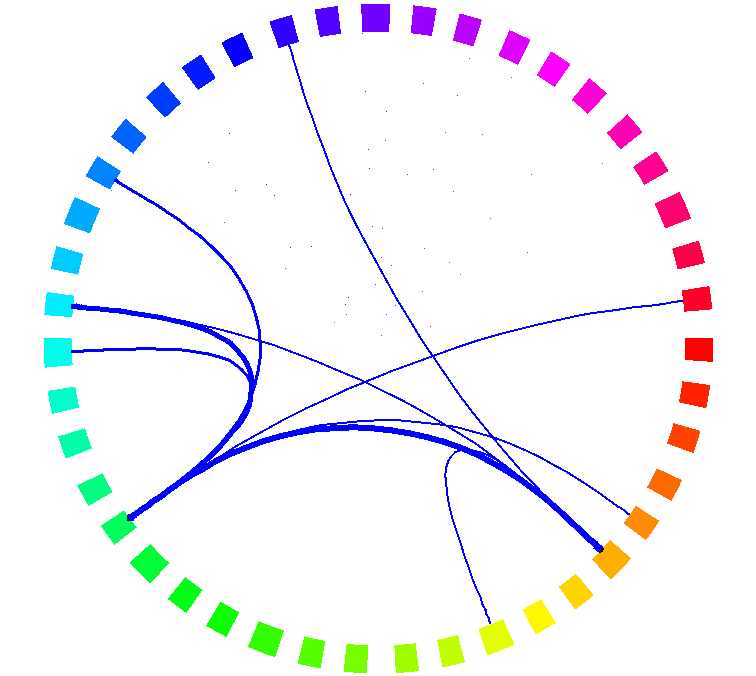}9
    \includegraphics[width=0.12\columnwidth]{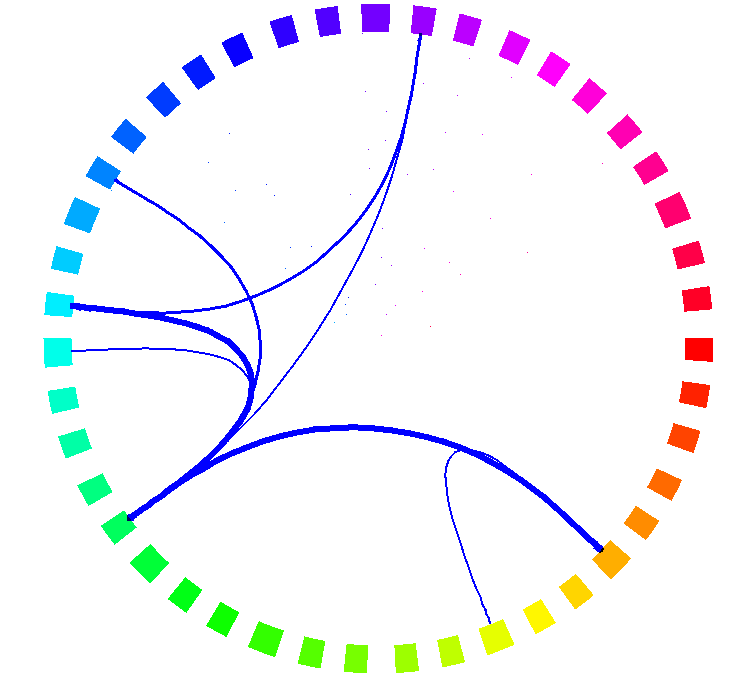}10
    \includegraphics[width=0.12\columnwidth]{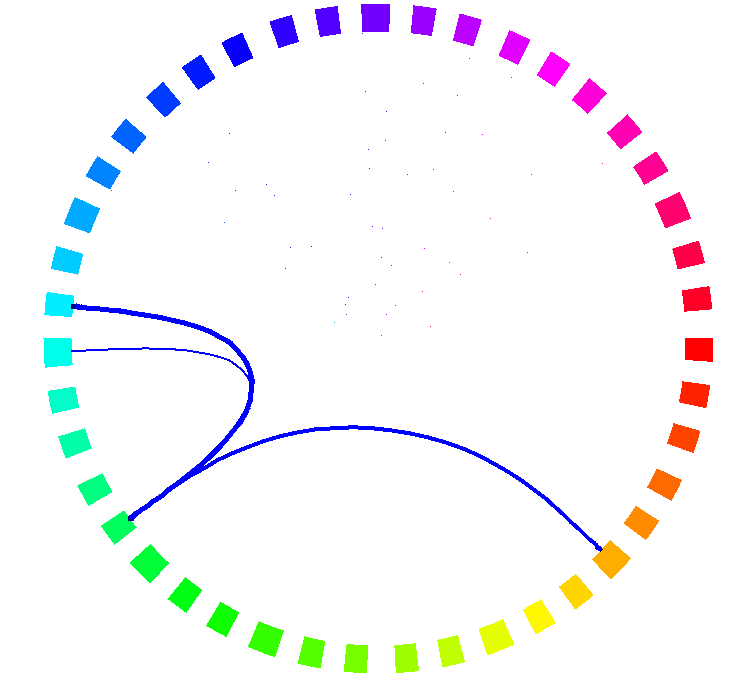}11
    \includegraphics[width=0.12\columnwidth]{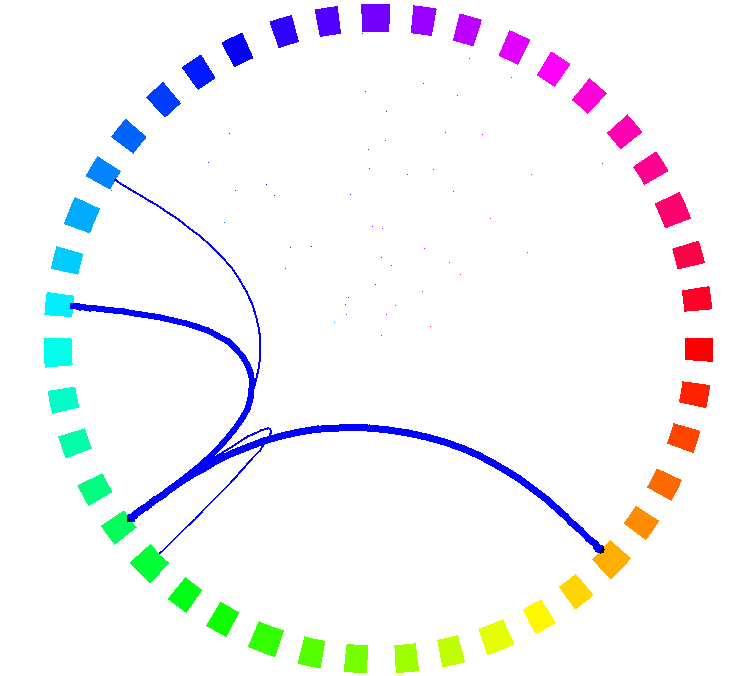}12
    \\
    &
    \includegraphics[width=0.12\columnwidth]{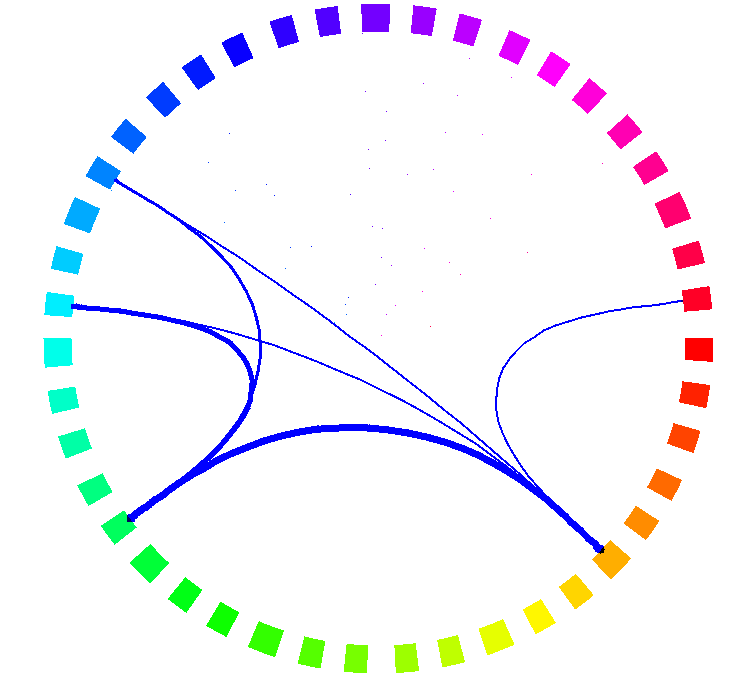}13
    \includegraphics[width=0.12\columnwidth]{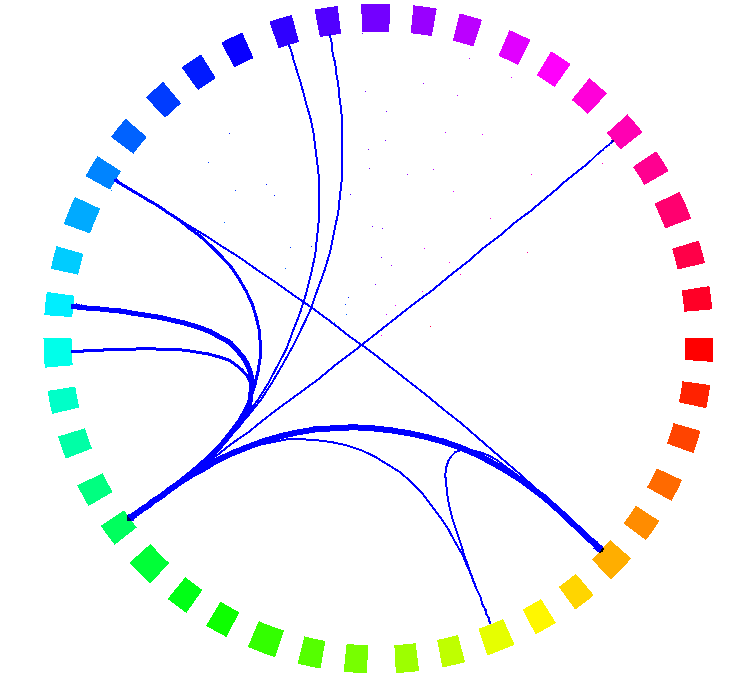}14
    \includegraphics[width=0.12\columnwidth]{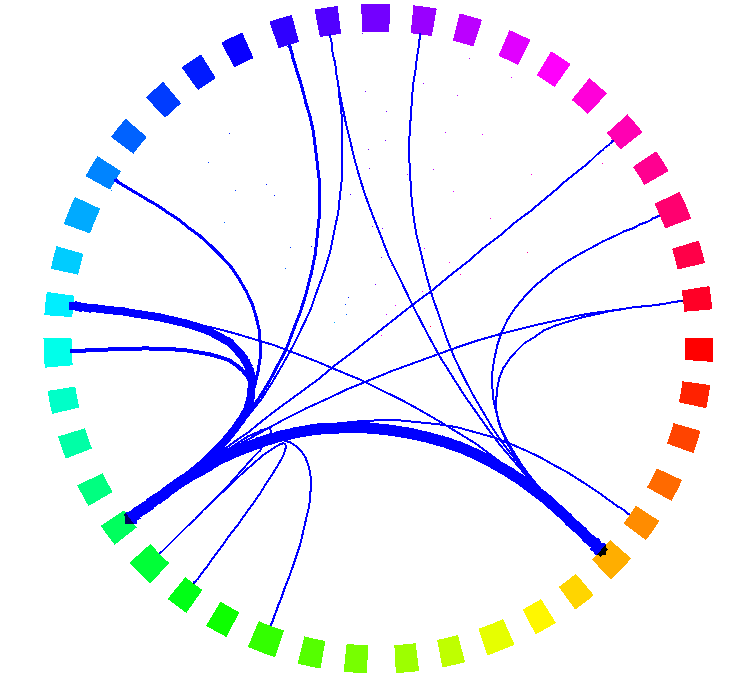}15
    \includegraphics[width=0.12\columnwidth]{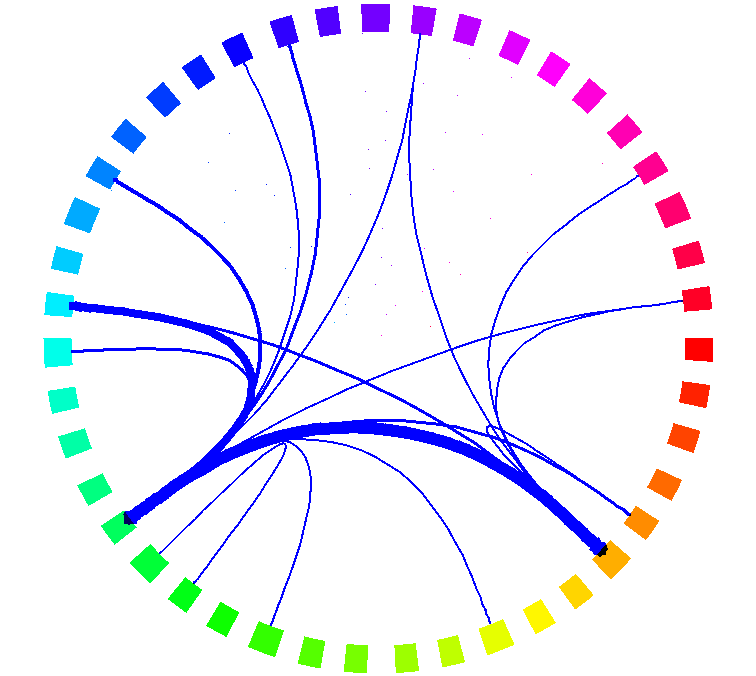}16
    \\
    \includegraphics[width=0.12\columnwidth]{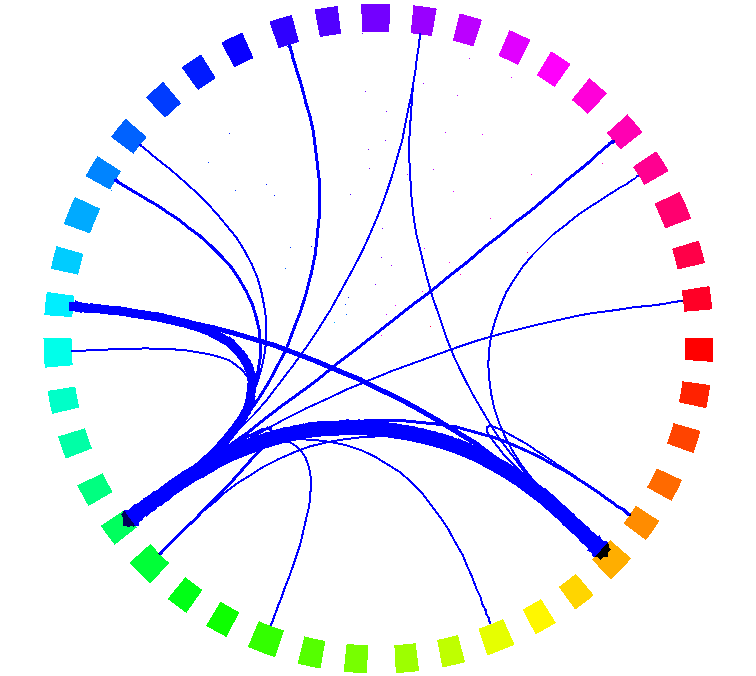}17
    \includegraphics[width=0.12\columnwidth]{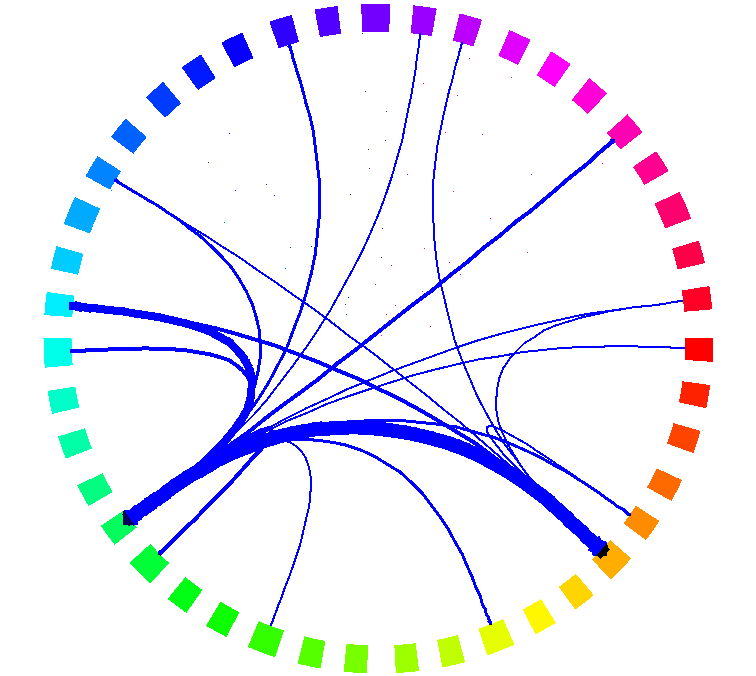}18
    \includegraphics[width=0.12\columnwidth]{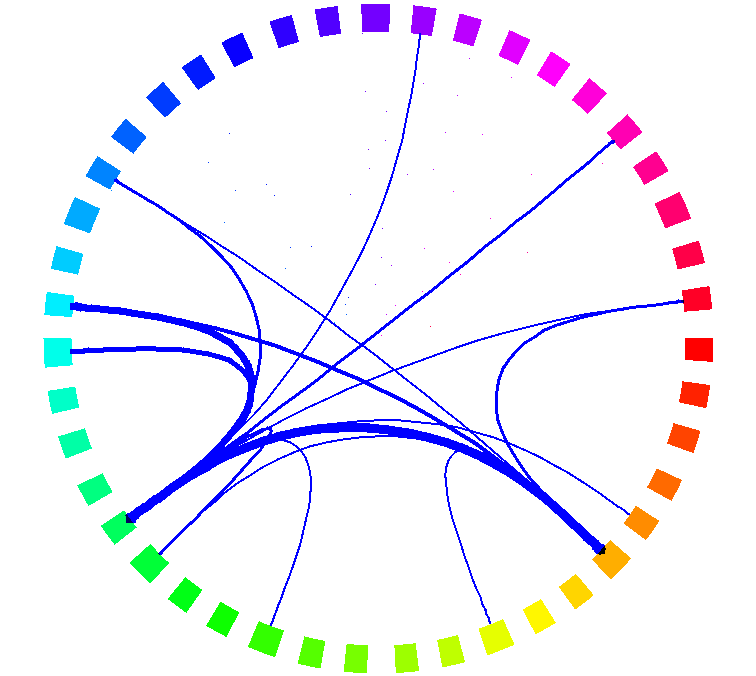}19
    &
    \includegraphics[width=0.12\columnwidth]{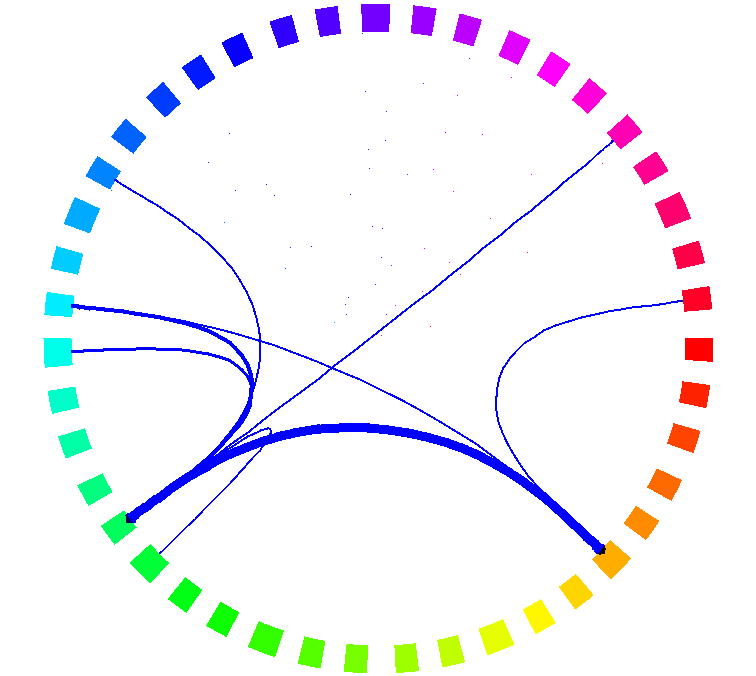}20
    \includegraphics[width=0.12\columnwidth]{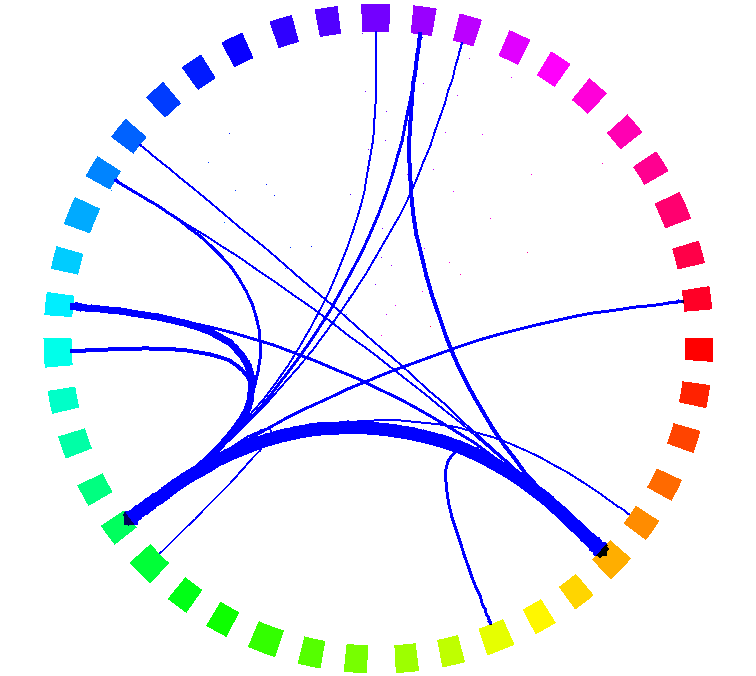}21
    \includegraphics[width=0.12\columnwidth]{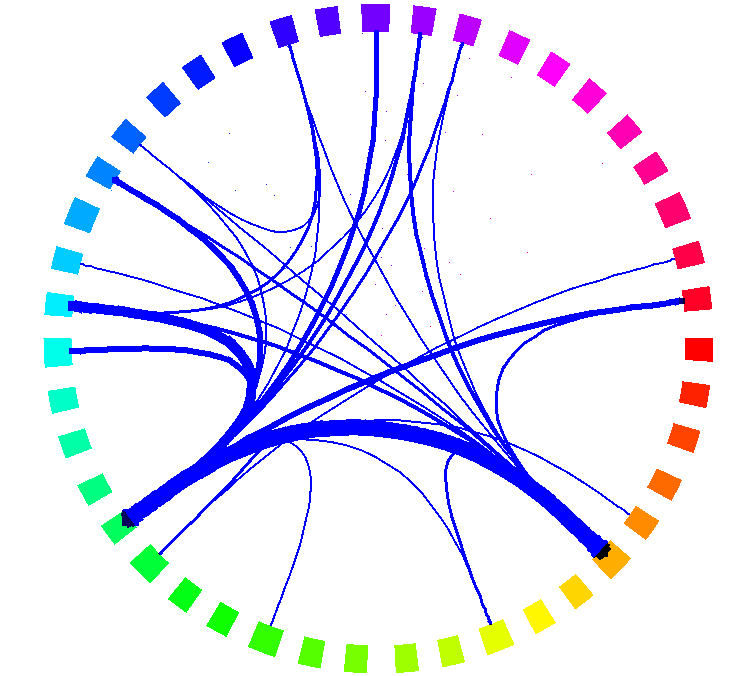}22
    \includegraphics[width=0.12\columnwidth]{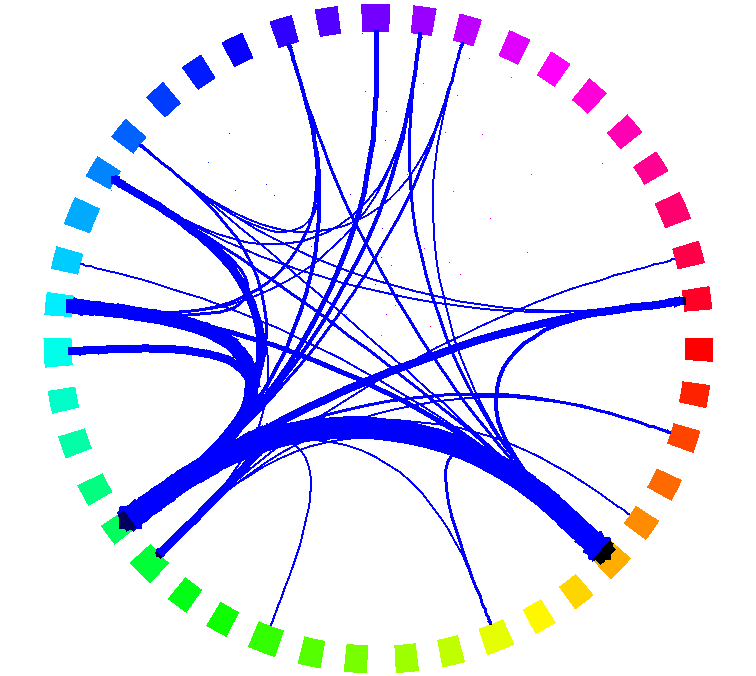}23
    \\
    \end{tabular}
    \caption{Temporally rewiring patterns of 43 gene ontology groups
    related to the cellular component, molecular function and biological
    process of \textit{Drosophila melanogaster}. The 4028 measured genes
    are grouped according to the 43 ontology group; The weight of an
    edge between two ontology groups counts the number of
    connections between genes across these two groups. The width of
    an edge in the visualization is proportional to its weight.
    For clarity, only edges with weights greater than 5 are
    displayed. Each ontology group
    is coded in clockwise order with a color shown in the legend on the upper left corner.}\label{fg:ontology}
\end{sidewaysfigure}

For this purpose, the 4028 genes are classified into 3 top level
gene ontology (GO) groups related to cellular component, molecular
function and biological process of \textit{Drosophila
melanogaster} according to Flybase. Then they are further divided
into 43 gene ontology groups which are the direct children of the
3 top level GO groups. The interaction between these ontology
groups evolving over time is shown in Fig.~\ref{fg:ontology}.

Through all stage of developmental process, genes belonging to
three ontology groups are most active, and they are related to
binding function, transcription regulator activity and organelle
function respectively. Particularly, the group of genes involved
in binding function play the central role as the hub of the
networks of interactions between ontology groups. Genes related to
transcriptional regulatory activity and organelles function show
persistent interaction with the group of genes related to binding
function. Other groups of genes that often interact with the
binding genes are those related to functions such as developmental
process, response to stimulus and biological regulation.

Large topological changes can also be observed from the temporal
rewiring patterns between these gene ontology groups. The most
diverse interactions between gene ontology groups occur at the
beginning of embryonic stage and near the end of adulthood stage.
In contrast, near the end of embryonic stage (time point 10), the
interactions between genes are largely restricted to those from 4
gene ontology groups: transcriptional regulator activity, enzyme
regulator activity, binding and organelle.

\subsection*{Transient Coherent Subgraphs}

During the development of \textit{Drosophila melanogaster}, there
may be sets of genes within which the interactions exhibit
correlated appearance and disappearance. These sets of genes and
their tight interactions form coherent subgraphs. Note that coherent
subgraphs can be different from the clusters in the summary graph.
While the former emphasizes the synchronous activation and
deactivation of edges over time, the later only concerns the
cumulative effect of the degree of interactions between genes.

To identify these coherent subgraphs, CODENSE~\citep{HuYanHua2005}
is applied to the inferred dynamic networks and 8 coherent
subgraphs are discovered~(Table~\ref{tb:CODENSE}). These coherent
subgraphs vary in size, with the smallest subgraph containing 27
genes and the largest subgraph containing 87 genes. The degree of
activity of these functional modules as measured by the clustering
coefficients of the subgraphs follows a stage-specific temporal
program. For instance, subgraph 3 and 4 are most active during the
adulthood stage, while subgraph 4 and 7 are most active during
embryonic and pupal stage respectively.

It is natural to ask whether the genes in a coherent subgraph are
enriched with certain functions. To show this, the functional
composition of these 6 subgraphs are also compared to the
background functional composition of the set of all 4028 genes.
Statistical test (two sample Kolmogorov-Smirnov test with
significance level 0.05) shows that 6 out of the 8 subgraphs are
significantly different from the background in term of their
functions. For instance, subgraph 6 is enriched with genes related
to binding (20.3\%), envelope (18.6\%), organelle part (16.9\%),
organelle (15.3\%) and antioxidant activity (5.1\%) ($p$-value
$<10^{-6}$). Furthermore, these genes reveal increased activity
near the end of the first 3 developmental stages. Another example
is subgraph 7 which is enrich with genes related to transporter
activity (11.1\%), reproductive process (11.1\%), multicellular
organismal process (11.1\%), organelle (7.4\%) and transcription
regulator activity (7.4\%). These genes peak in their activity
near the end of the pupal stage. These results suggest that
different gene functional modules follow very different temporal
programs.

\begin{table}
  \centering
  \caption{CODENSE is used to discover coherent subgraphs
  across the 23 temporally rewiring networks. Six such subgraphs
  are discovered. The number of genes within each subgraph are 69,
  58, 29, 87, 64, 59, 27 and 63 respectively.
  In each cell of the table the evolution of the clustering coefficient for each subgraph is plotted, and
  the corresponding configurations of the subgraph at different time points are also illustrated.
  }\label{tb:CODENSE}
  \scalebox{1}{
  \begin{tabular}{|c|c|c|c}
    \hline
    \hline
    \includegraphics[width=0.20\columnwidth]{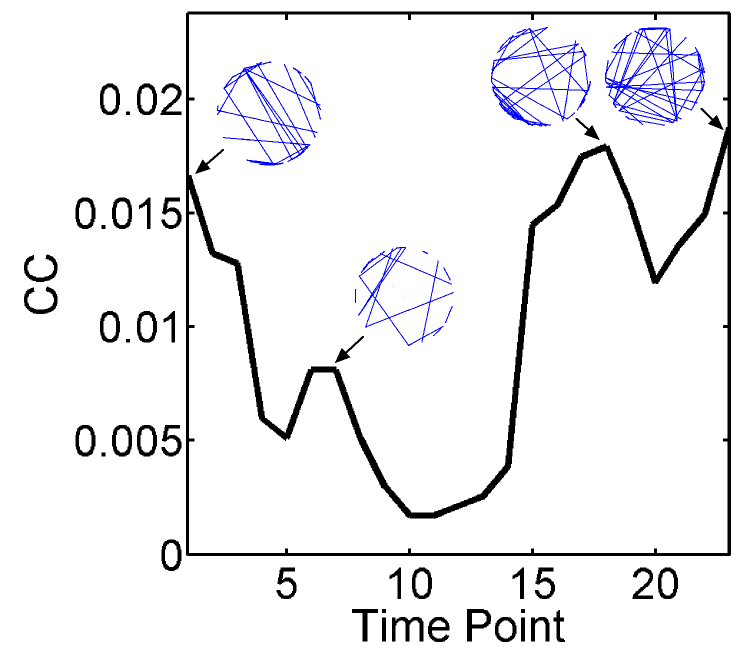}1
    &
    \includegraphics[width=0.20\columnwidth]{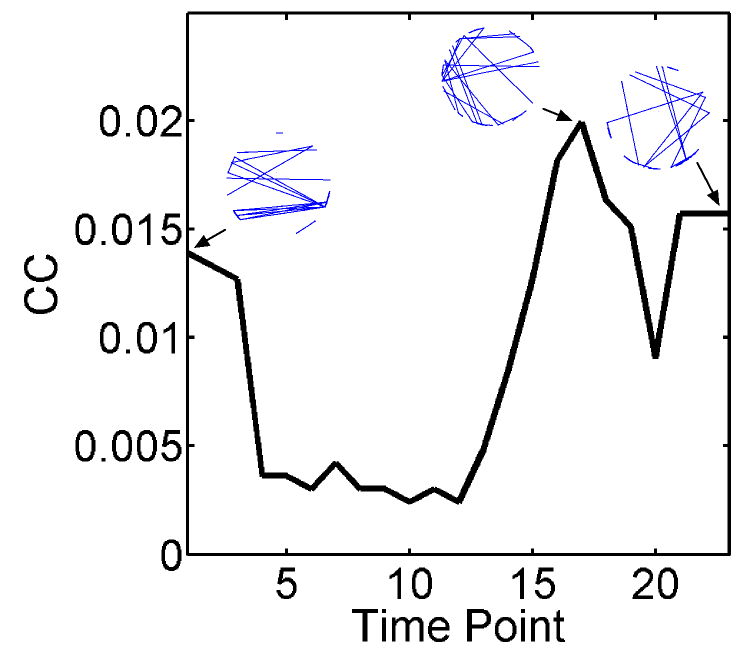}2
    &
    \includegraphics[width=0.20\columnwidth]{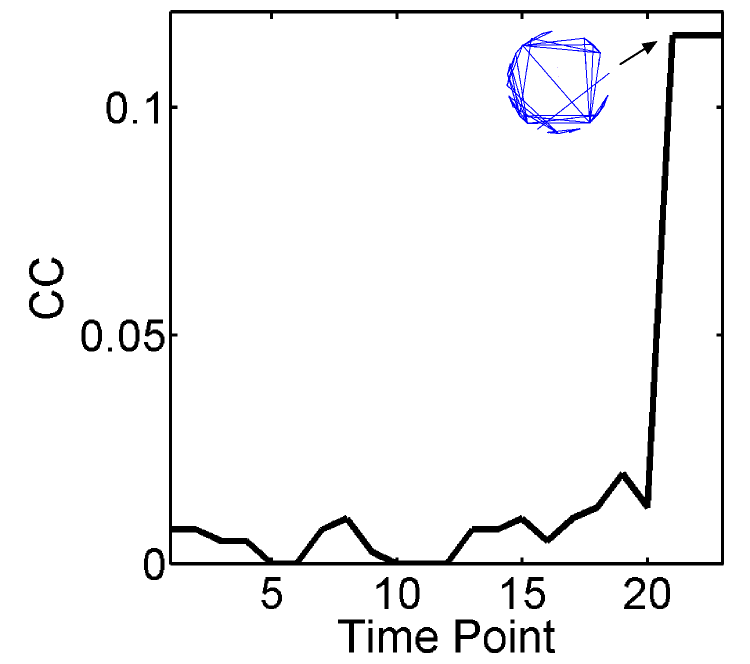}3
    &
    \includegraphics[width=0.20\columnwidth]{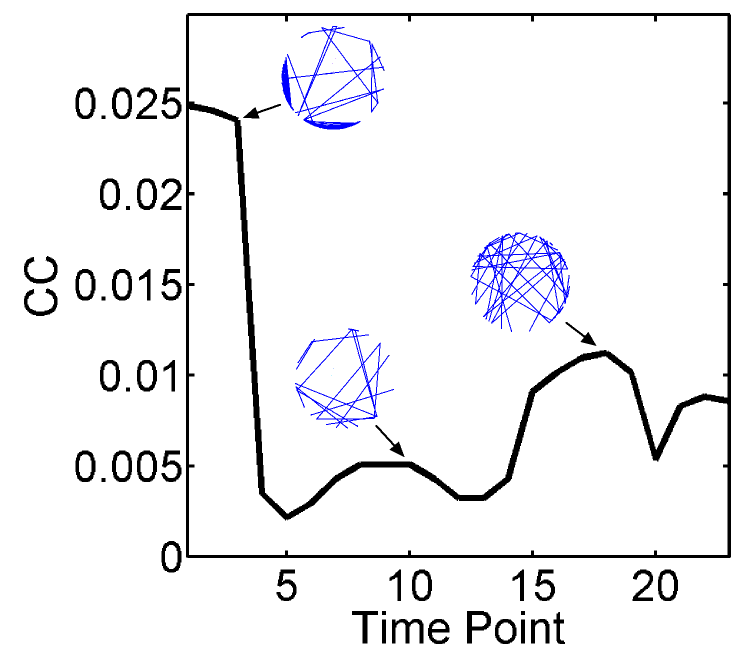}4
    \\
    \includegraphics[width=0.20\columnwidth]{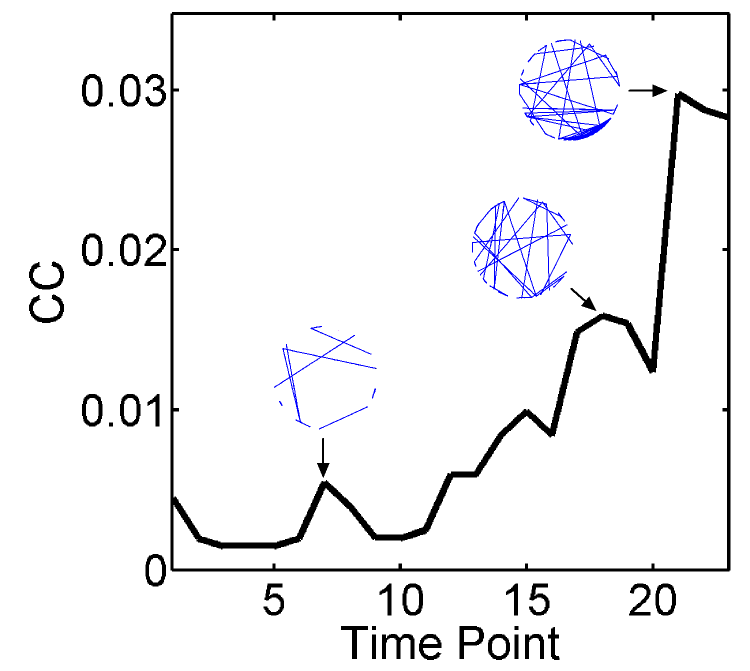}5
    &
    \includegraphics[width=0.20\columnwidth]{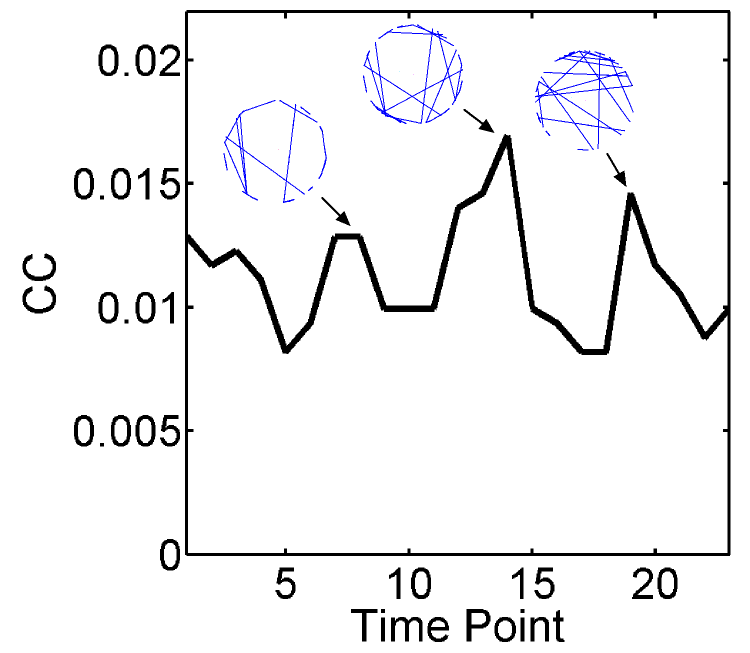}6
    &
    \includegraphics[width=0.20\columnwidth]{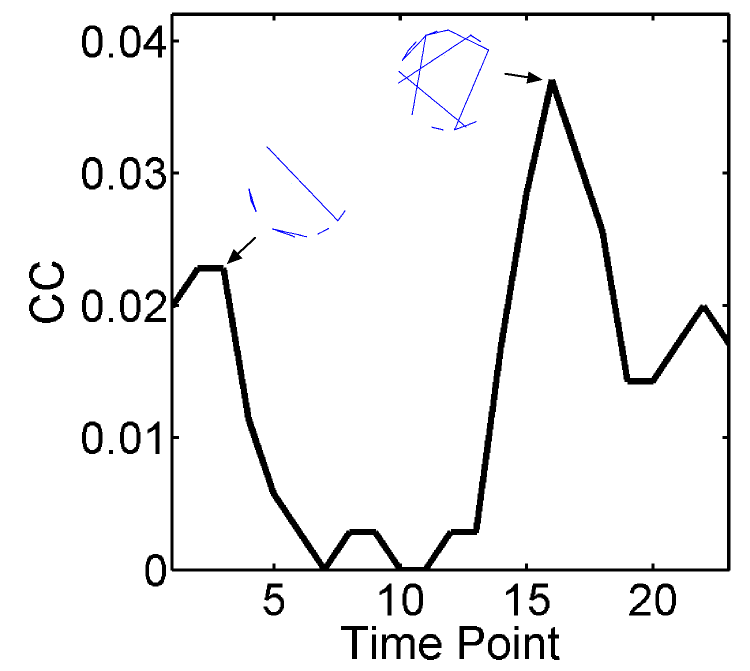}7
    &
    \includegraphics[width=0.20\columnwidth]{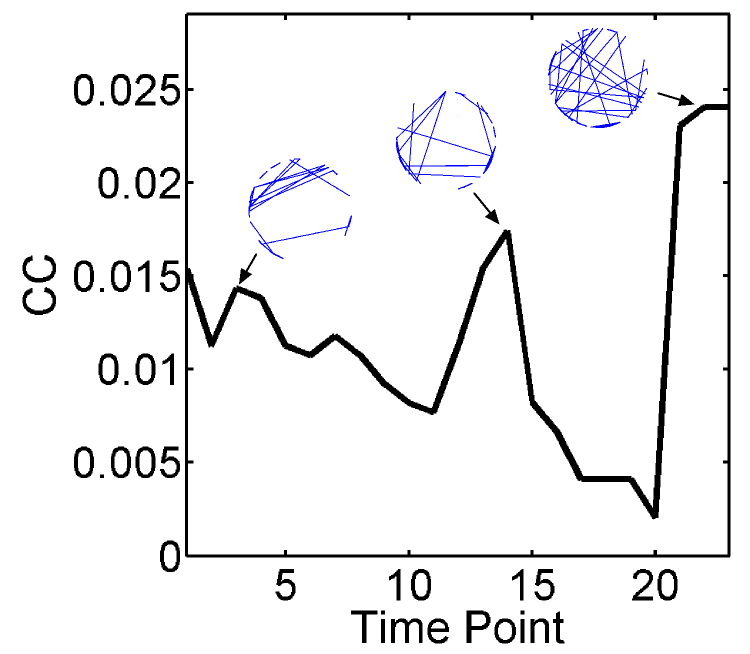}8
    \\
    \hline
    \hline
  \end{tabular}}
\end{table}

\subsection*{Dynamics of Known Gene Interactions}

Different gene interactions may following distinctive temporal
programs of activation, appearing and disappearing at different
time point during the life cycle of \textit{Drosophila
melanogaster}. In turn the transient nature of the interactions
implies that the evidence supporting the presence of an
interaction between two genes may not be present in all microarray
experiments conducted during different developmental stages of the
organism. Therefore, pooling all microarray measurements and
inferring a single static network can undermine the inference
process rather than helping it. This problem can be overcome by
learning dynamic networks which recover transient interactions
that are supported by correct subsets of experiments.

To show the advantage of dynamic networks over a static network,
the recovered interaction by these two types of networks are
compared against a list of 1143 known undirected gene interactions
hosted in Flybase. The dynamic networks recover 96 of these known
gene interactions while the static network only recovers 64. That
is dynamic networks recover 50\% more known gene interactions than
the static network. Furthermore, the static network provides no
information on when a gene interaction starts or ends. In
contrast, the dynamic networks pinpoint the temporal on-and-off
sequence for each recovered gene interaction.

\begin{sidewaysfigure}
  \includegraphics[bb= 0 0 962 692,width=0.48\columnwidth]{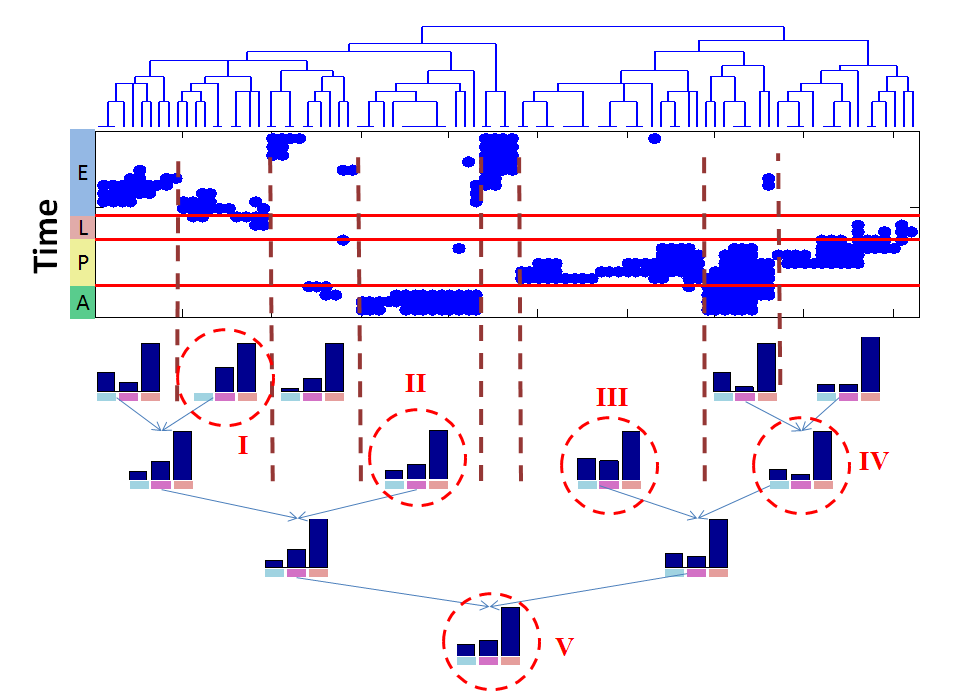}(a)
  \includegraphics[bb= 0 0 996 8 66,width=0.48\columnwidth]{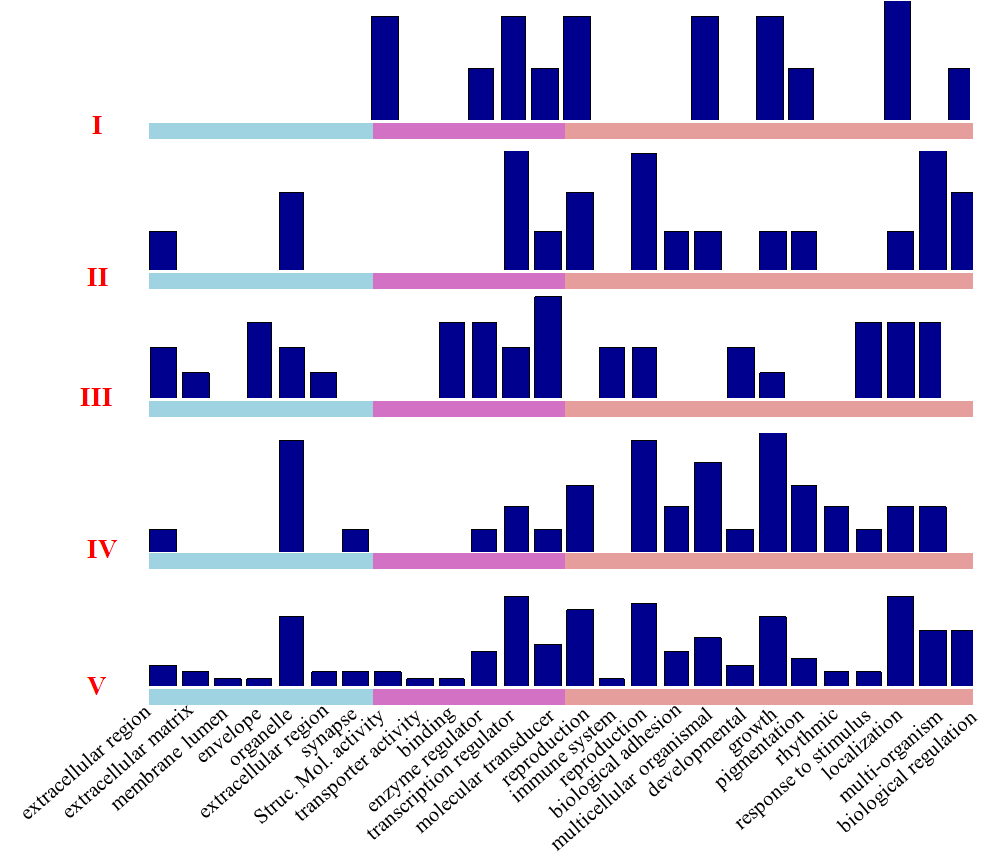}(b)
  \caption{Ninety six known gene interactions are recovered by the inferred dynamic networks.
  The onset and duration of these interactions follow different temporal patterns.
  The activation of each gene interaction over time
  is represented as one column in the upper part of (a);
  within each column, if a gene interaction is active a
  blue dot is drawn otherwise the space is left blank. These
  temporal sequences of activation can also be viewed as a sequence of zeros and
  ones where zero means not active and one means active. Based on
  these sequences of zeros and ones, hierarchical clustering is
  performed on the gene interactions. The clustering results are
  displayed on the top of the activation patterns. Note that the
  gene interactions have also been ordered according to the
  hierarchical clustering results. Therefore block of gene
  interactions with similar activation patterns can be seen. To investigate
  how the activation patterns of gene interactions are related to their biological function,
  several clusters of gene interactions are highlighted as cluster I, II, III
  IV and V. The 43 gene ontology terms are used as the bins to build histograms
  to investigate the functional composition of each cluster. The histograms are enlarged and
  shown in (b). The histogram bins are ordered according to the
  gene ontology groups shown in the horizontal axis. The colors of
  the axis indicate the top 3 level of the gene ontology group
  related to cellular component (cyan), molecular function (purple) and biological
  process (pink).
  }
  \label{fg:knowngene}
\end{sidewaysfigure}

To investigate whether the gene interactions with different
activation pattern are related to their functional difference,
hierarchical clustering is performed on these set of recovered
gene interactions based on their activation patterns
(Fig.~\ref{fg:knowngene}). It can be seen that all these
interactions are transient and very specific to certain stage of
the life cycle of \textit{Drosophila melanogaster}. Furthermore,
gene interactions with similar activation pattern tends to recruit
genes with similar functions. To demonstrate this, histogram
analysis is performed base on the gene ontology groups of the
genes involved in interactions. In Fig~\ref{fg:knowngene}, five
clusters in different level of the cluster hierarchy are
highlighted and their histograms show that functionally different
gene interactions tend to activation in very different temporal
sequence. For instance, gene interactions in cluster I activates
near the boundary of embryonic and larval stage, and in these
interactions, no genes are related to the cellular component
function of the organism; cluster II activates near the end of
pupal stage and is enriched by genes related to the cellular
component function.

\section*{Conclusion}

Numerous algorithms have been developed for inferring biological
networks from high throughput experimental data, such as microarray
profiles ~\citep{Segal_nature,Dobra_west04,ong}, ChIP-chip genome
localization data~\citep{young2002,Ziv_Nat03,young2004}, and
protein-protein interaction (PPI)
data~\citep{UetzPPI00,rothberg2003,ideker2004,causier2004}, based on
formalisms such as graph mining~\citep{Shamir_PNAS04}, Bayesian
networks~\citep{cowell}, and dynamic Bayesian
networks~\citep{Kanazawa95}. However, most of this vast literature
focused on modeling a static network or time-invariant networks
~\citep{Friedman2000}, and much less has been done towards modeling
the dynamic processes underlying networks that are topologically
rewiring and semantically evolving over time, and on developing
inference and learning techniques for recovering unobserved network
topologies from observed attributes of entities (e.g., genes and
proteins) constituting the network. The Tesla program presented here
represents the first successful and practical tool for genome-wide
reverse engineering the network dynamics based on the gene
expression and ontology data. This method allows us to recover the
wiring pattern of the genetic networks over a time series of
arbitrary resolution. The recovered networks with this unprecedented
resolution chart the onset and duration of many gene interactions
which are missed by typical static network analysis. We expect
collections of complex, high-dimensional, and feature- rich data
from complex dynamic biological processes such as cancer
progression, immune response, and developmental processes to
continue to grow, given the rapid expansion of categorization and
characterization of biological samples, and the improved data
collection technologies. Thus we believe our new method is a timely
contribution that can narrow the gap between imminent methodological
needs and the available data, and offer deeper understanding of the
mechanisms and processes underlying biological networks.

\section*{Material and Methods}

\subsection*{Microarray Data}

We used the microarray data collected by~\cite{Arbeitmanetal2002}
in their study of the gene expression patterns during the life
cycle of \textit{Drosophila melanogaster}. Approximately 9,700
\textit{Drosophila} cDNA elements representing 5,081 different
genes were used to construct the 2-color spotted cDNA microarrays.
The genes analyzed in this paper consists of a subset of 4,028
sequence-verified, unique genes. Experimental samples were
measured at 66 different time points spanning the embryonic,
larval, pupal and adulthood period. Each hybridization is a
comparison of one sample to a common reference sample made from
pooled mRNA representing all stages of the life cycle.
Normalization is performed so that the dye dependent intenstive
response is removed and the average ratio of signals from the
experimental and reference sample equals one. The final expression
value is the log ratio of signals.

\subsection*{Missing Value Imputation}

Missing values are imputed in the same manner
as~\cite{ZhaSerDou2006}. This is based on the assumption that gene
expression values change smoothly over time. If there is a missing
value, a simple linear interpolation using values from adjacent
time points is used, i.e. the value of the missed time point is
set to the mean of its two neighbors. When the missing point is a
start or a end point, it is simply filled with the value of its
nearest neighbor.

\subsection*{Expression Value Binarization}
The expression values are quantized into binary numbers using
thresholds specific to each gene in the same manner
as~\cite{ZhaSerDou2006}. For each gene, the expression values are
first sorted; then the top two extreme values in either end of the
sorted list are discarded; last the median of the remaining values
are used as the threshold above which the value is binarized as 1
and 0 otherwise. Here 1 means the expression of a gene is
up-regulated, and 0 means down-regulated.

\subsection*{Network Inference model}

In this paper, we used a new  approach for recovering
time-evolving networks on \emph{fixed} set of genes from time
series of gene expression measurements using {\it temporally
smoothed $L_1$-regularized logistic regression}, or in short, TLR
(for temporal LR). This TLR can be formulated and solved using
existing efficient convex optimization techniques which makes it
readily scalable to learning evolving graphs on a genome scale
over few thousands of genes.

For each time epoch we assumed that we observe binary gene
activation patterns, which we obtained from the continuous
micro-array measurements as described above. At each epoch we
represent the regulatory network using a Markov random field
define as follows:

Let $G^t = (V, E^t)$ be the graph structure at time epoch $t$ with
vertex set $V$ of size $|V|$ = p and edge set $E^t$. Let
$\{X^{t}_{1:N_t}\}$ be a set of i.i.d \emph{binary} random
variables associated with the vertices of the graph. Let the joint
probability of the random variables be given by the Ising model as
follows:
\begin{eqnarray}
P(\mathbf{x^t_d}|\Theta^t) &=& {\rm{exp}} \Bigg( \sum_{i\in
V}{\theta^t_{ii} x_{d,i}^t} + \sum_{(i,j) \in
E^t}{\theta^t_{ij}x^t_{d,i} x^t_{d,j} - A(\Theta^t)}\Bigg),
\label{eq:emision}
\end{eqnarray}
where the parameters $\{\theta^t_{ij}\}_{(i,j) \in E^t}$ capture
the correlation (regulation strength) between genes $X^t_i$, and
$X^t_j$. $A(\Theta^t)$ is the log normalizing constant of the
distribution. Given a set of $\{X^{t}_{1:N_t}\}$ i.i.d samples
drawn from $P(X^t_d|\Theta^t)$ at each time step, the goal is to
estimate the structure of the graph, i.e. to estimate
$\{\hat{E^t}\}_{t=1}^T$. We cast this problem as a regularized
estimation problem of the time-varying parameters of the graph as
follows: \vspace{-0.2cm}

\begin{eqnarray}
\hat{\Theta^1}, \ldots, \hat{\Theta^T} &=& \arg\min_{\Theta^1,
\ldots, \Theta^T} \,\,\, \sum_{t=1}^{T}{nLL(\Theta^t)} \,\, + \,\,
R(\Theta^1 \ldots \Theta^T,\lambda), \label{eq:learningProblem}
\end{eqnarray}

where, $nLL(\cdot)$ is the exact (or approximate surrogate) of the
negative Log Likelihood $R(\cdot)$  is a regularization term, and
$\lambda$ is the regularization parameter(s). We assume that the
graph is \emph{sparse} and evolves \emph{smoothly} over time, and
we would like to pick a regularization function $R(\cdot)$ that
results in a sparse and smooth graphs. The structure of the graphs
can then be recovered from the non-zero parameters which are
isomorphic to the edge set of the graphs.

When $T=1$, the problem in Eq. (\ref{eq:learningProblem})
degenerates to the static case:
\begin{eqnarray}
\hat{\Theta} &=& \arg\min_\Theta \, nLL(\Theta) +
R(\Theta,\lambda), \label{eq:static}
\end{eqnarray}
and thus one needs only to use a regularization function,
$R(\cdot)$, that enforces sparsity. Several approaches were
proposed in the literature \citep{Martin,Daphne,htERGM}. this
problem has been addressed by choosing $R=L_1$-penalty, however,
they differ in the way they \emph{approximate} the first term in
Eq. (\ref{eq:static}) which is intractable in general due to the
existence of the log partition function, $A(\Theta)$
\citep{Martin,Daphne,htERGM}. In \cite{Martin} a pseudo-likelihood
approach was used. The pseudo-likelihood, $\hat{P}(X_d|\Theta) =
\prod_{i=1}^P{P\big(x_{d,i}|\mathbf{x_{d,N(i)}}\big)}$, where
$N(i)$ is the Markov blanket of node $i$, i.e., the neighboring
nodes of node $i$. In the binary pairwise-MRF, this local
likelihood has a logistic-regression form. Thus the learning
problem in (\ref{eq:static}) degenerates to solving $P$
$l_1$-regularized logistic regression problems resulting from
regressing each individual variable on all the other variables in
the graph. More specifically, the learning problem for node $i$ is
given by:
\begin{eqnarray}
\hat{\mathbf{\theta}_i}&=& \arg\min_{\theta_i} \frac{1}{N} \sum^{N}_{d=1}{\log\,P\big(x_{d,i}|\mathbf{x_{d,-i}},\theta_i\big)}+ \lambda_1 \parallel \theta_{-i}\parallel_{1} \nonumber\\
&=& \arg\min_{\mathbf{\theta_i}} \frac{1}{N}
\sum^{N}_{d=1}{\big[\mathrm{log}(1+\mathrm{exp}(\mathbf{\theta_i}
\mathbf{x_{d,-i})}) - x_{d,i}\mathbf{\theta_i} \mathbf{x_{d,-i}}
\big]}+\lambda_1
\parallel \mathbf{\theta_{-i}}\parallel_{1},  \label{eq:LRP}
\end{eqnarray}
where, $\mathbf{\theta_i} = (\theta_{i1}, \ldots,\theta_{iP})$ are
the parameters of the $L_1$-logistic regression,
$\mathbf{x_{d,-i}}$ denotes the set of all variables with
$x_{d,i}$ replaced by 1, and $\theta_{-i}$ denotes the vector
$\theta_{i}$ with the component $\theta_{ii}$ removed (i.e. the
intercept is not penalized). The estimated set of neighbors is
given by: $\hat{N}(i) = \{j : \theta_{ij} \neq 0\}$. The set of
edges $E$ is then defined as either a union or an intersection of
neighborhood sets $\{N(i)\}_{i \in V}$ of all the vertices.
\cite{Martin} showed that both definitions would converge to the
true structure asymptotically.

To extend the above approach to the dynamic setting,  we return to
the original learning problem in Eq. (\ref{eq:learningProblem})
and approach it using the techniques presented above. We use the
negative pseudo-loglikelihood as a surrogate for the intractable
$nLL(\cdot)$ at each time epoch. To constrain the multiple
time-specific regression problems each of which takes the form of
Eq. (\ref{eq:LRP}) so that graphs are evolving in a smooth
fashion, that is, not dramatically rewiring over time, we penalize
the difference between the regression coefficient vectors
corresponding to the same node, say $i$, at the two adjacent time
steps. This can be done by introducing a regularization term
$\parallel \theta_{i}^t - \theta_{i}^{t-1}\parallel_{1}^1$ for
each node at each time.  Then, to also enforce sparsity over the
graphs learnt at each epoch, in addition to the smoothness between
their evolution across epochs, we use the standard $L_1$ penalty
over each $\theta_{i}^t$.  These choices will decouple the
learning problem in (\ref{eq:learningProblem}) into a set of $P$
separate \emph{smoothed} $L_1$-regularized logistic regression
problems, one for each variable. Putting everything together, for
each node $i$ in the graph, we solve the following problem for :
\begin{eqnarray}
\hat{\theta}_i^1, \ldots, \hat{\theta}_i^T &=&
\arg\min_{\theta_i^1, \ldots, \theta_i^T}
\sum^{T}_{t=1}{l_{avg}(\theta_i^t)} +\lambda_1
\sum^{T}_{t=1}{\parallel \theta_{-i}^t\parallel_{1}} + \lambda_2
\sum^{T}_{t=2}{ \parallel \theta_{i}^t -
\theta_{i}^{t-1}\parallel_{1}^1}, \,\,\, \label{eq:dynamic}
\end{eqnarray}
where
\begin{eqnarray}
\,\,l_{avg}(\mathbf{\theta_i^t}) &=& \frac{1}{N^t}
\sum^{N^t}_{d=1}{\log\,P\big(x^t_{d,i}|\mathbf{x^t_{d,-i}},\mathbf{\theta^t_i}\big)}
\nonumber \\
&=& \frac{1}{N^t}
\sum^{N^t}_{d=1}{\big[\mathrm{log}(1+\mathrm{exp}(\mathbf{\theta^t_i}
\mathbf{x^t_{d,-i}})) - x^t_{d,i}\mathbf{\theta^t_i}
\mathbf{x^t_{d,-i}} \big]}.
\end{eqnarray}

The problem  in Eq. (\ref{eq:dynamic}) is a convex optimization
problem with a non-smooth $L_1$ functions. Therefore, we solve the
following equivalent problem instead by introducing new auxiliary
variables, $\mathbf{u^t_i}$ and $\mathbf{v^t_i}$ (the case for
$q=2$ is handled similarly):
\begin{eqnarray}
&&\min_{\theta_i^1, \ldots, \theta_i^T,u_i^1, \ldots, u_i^T,v_i^2,
\ldots,v_i^T} \sum^{T}_{t=1}{l_{avg}(\theta_i^t)} +\lambda_1
\sum^{T}_{t=1}\mathbf{1}^T{\mathbf{u^t_i}} +
\lambda_2 \sum^{T}_{t=2}{\mathbf{1}^T \mathbf{v^t_i}} \label{eq:optimization}\\
&&\mathrm{subject \, to} \quad -u^t_{i,j} \leq \theta^t_{i,j} \leq u^t_{i,j}, \,\, t=1,\ldots,T, j=1,\ldots,i-1,i+1,\ldots,P, \nonumber\\
&&\mathrm{subject \, to} \quad -v^t_{i,j} \leq \theta^t_{i,j} -
\theta^{t-1}_{i,j} \leq v^t_{i,j}, \,\, t=2,\ldots,T, j=1,\ldots,
P , \nonumber
\end{eqnarray}

where $\mathbf{1}$ denotes a vector with all components set to 1,
so $\mathbf{1}^T\mathbf{u^t_i}$ is the sum of the components of
$\mathbf{u^t_i}$. To see the equivalence of the problem in Eq.
(\ref{eq:optimization}) with the one in Eq. (\ref{eq:dynamic}, we
note that at the optimal point of Eq. (\ref{eq:optimization}), we
must have $u^t_{i,j}=|\theta^t_{i,j}|$, and similarly
$v^t_{i,j}=|\theta^t_{i,j}- \theta^{t-1}_{i,j}|$, in which case
the objectives in Eq. (\ref{eq:optimization}) and Eq.
(\ref{eq:dynamic}) are the same (a similar solution has been
applied to solving $L_1$-regularized logistic regression in
\cite{boyedL1}).  The problem in Eq. (\ref{eq:optimization}) is a
convex optimization problem, with now a smooth objective, and
linear constraint functions, so it can be solved by standard
convex optimization methods, such as interior point methods, and
high quality solvers can directly handle the problem in Eq.
(\ref{eq:optimization}) efficiently for medium to large scale (up
to few thousands of nodes). In this paper, we used the CVX
optimization
package\footnote{\url{http://stanford.edu/$\sim$boyd/cvx}}.

\subsection*{Gene Ontology}

Gene ontology (GO) data is obtained from
Flybase\footnote{\url{http://flybase.bio.indiana.edu}}. There are
altogether 25,072 GO terms with 3 name spaces: molecular function,
biological process and cellular component. The GO terms are
organized into a hierarchical structure, and 16,189 GO terms are
leaf nodes of the hierarchy. The 4,028 genes are assigned to one
or more leaf nodes. When we estimate the networks for a target
gene, we restrict the network inference to those genes that share
common GO terms with the target gene. This operation restricts the
network inference to a biologically plausible set of genes and it
also drastically reduces the computation time.

\subsection*{Mining Frequent Coherent Subgraphs}

CODENSE is an algorithm for efficiently mining frequent coherent
dense subgraphs across a large number of massive
graphs~\citep{HuYanHua2005}. In the context of dynamic networks,
it is adapted to discover functionally coherent gene modules
across the temporal snapshots of the networks. In such a module,
the interactions between component genes follow a similar pattern
of onset and duration.

CODENSE takes as inputs a summary graph and a support vector for
each edge. A node in the summary graph represents a gene; an edge
represents a gene interaction and its weight is the number of
times that this interaction occurs over time. The summary graph is
used to guide the search for dense subgraphs in CODENSE. A support
vector records the exact temporal sequence of on-and-off of a gene
interaction. Using the support vectors, CODENSE finds a set of
genes whose interactions follow a similar path of evolution.

\section*{Acknowledge}

Le Song is supported by Lane Fellowship for computational biology.

\clearpage
\bibliographystyle{natmlapa}
\bibliography{shortBib}
\end{document}